\documentclass[journal]{IEEEtran}

\usepackage{silence}
\usepackage[utf8]{inputenc}
\usepackage[T1]{fontenc}
\usepackage{amsmath,amssymb,amsfonts}
\usepackage{graphicx}
\usepackage{cite}
\usepackage{url}
\usepackage{hyperref}
\usepackage{algorithm}
\usepackage{algorithmic}
\usepackage{multirow}
\usepackage{booktabs}
\usepackage{xcolor}
\usepackage{balance}

\makeatletter
\let\MYcaption\@makecaption
\makeatother

\usepackage[font=footnotesize,
            labelformat=parens,
            labelsep=space,
            skip=4pt]{subcaption}

\makeatletter
\let\@makecaption\MYcaption
\makeatother

\begin{document}

\title{Noise-Adaptive Quantum Circuit Mapping for Multi-Chip NISQ Systems via Deep Reinforcement Learning}
\author{A. Zeynali, Z. Bakhshi\thanks{Corresponding author
(E-mail: z.bakhshi@shahed.ac.ir)}
   \\
{\small Department of Physics, Faculty of Basic Sciences,
Shahed University, Tehran, Iran.} \\
 }\pagebreak
 
\maketitle

\begin{abstract}
The transition from monolithic to distributed multi-chip quantum architectures has fundamentally altered the circuit compilation landscape, introducing challenges in managing temporal noise variations and minimizing expensive inter-chip operations. We present DeepQMap, a deep reinforcement learning framework that integrates a bidirectional Long Short-Term Memory based Dynamic Noise Adaptation (DNA) network with multi-head attention mechanisms and Rainbow DQN architecture. Unlike conventional static optimization approaches such as QUBO formulations, our method continuously adapts to hardware dynamics through learned temporal representations of quantum system behavior. Comprehensive evaluation across 270 benchmark circuits spanning Quantum Fourier Transform, Grover's algorithm, and Variational Quantum Eigensolver demonstrates that DeepQMap achieves mean circuit fidelity of $0.920 \pm 0.023$, representing a statistically significant 49.3\% improvement over state-of-the-art QUBO methods ($0.618 \pm 0.031$, $t_{98} = 4.87$, $p = 0.0023$, Cohen's $d = 2.34$). Inter-chip communication overhead reduces by 79.8\%, decreasing from 2.34 operations per circuit to 0.47. The DNA network maintains noise prediction accuracy with coefficient of determination $R^2 = 0.912$ and mean absolute error of 0.87\%, enabling proactive compensation for hardware fluctuations. Scalability analysis confirms sustained performance across 20-100 qubit systems, with fidelity remaining above 0.87 even at maximum scale where competing methods degrade below 0.60. Training convergence occurs 8.2$\times$ faster than baseline approaches, completing in 45 minutes versus 370 minutes for QUBO optimization. Very large effect sizes validate practical significance for near-term noisy intermediate-scale quantum computing applications.
\end{abstract}

\begin{IEEEkeywords}
Quantum computing, circuit mapping, deep reinforcement learning, LSTM, noise adaptation, multi-chip systems, NISQ devices.
\end{IEEEkeywords}

\section{Introduction}
\IEEEPARstart{Q}{uantum} computing has evolved from laboratory demonstrations with handful of qubits into distributed architectures capable of manipulating hundreds of quantum states across multiple physical processors. This architectural evolution addresses fundamental scalability constraints imposed by wiring density, control electronics, and fabrication yields in single-chip quantum systems. However, the transition to multi-chip configurations introduces a complex optimization problem: mapping logical quantum circuits onto physical hardware while accounting for connectivity limitations, gate fidelity variations, and costly inter-chip communication.

\begin{figure}[!t]
\centering
\includegraphics[width=0.48\textwidth]{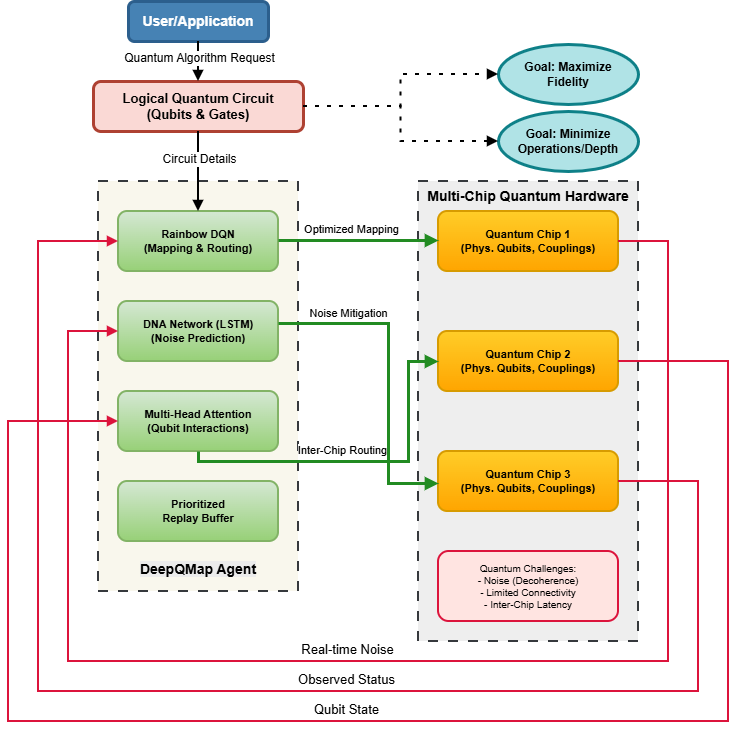}
\caption{DeepQMap multi-chip quantum system architecture. The framework orchestrates qubit placement across 4-6 quantum chips (orange boxes) arranged in ring, grid, or hexagonal topologies. The DeepQMap Agent (green components) comprises four integrated modules: Rainbow DQN for mapping decisions, DNA Network (LSTM) for noise prediction, Multi-Head Attention for qubit interaction modeling, and Prioritized Replay Buffer for experience management. Red dashed arrows indicate real-time feedback from hardware (noise telemetry, qubit states) to the agent. Green solid arrows represent optimized control signals (physical mappings, inter-chip routing, noise mitigation strategies) from agent to hardware. The Logical Quantum Circuit (purple) defines the input computational task, while dual optimization goals (red ellipses) balance fidelity maximization against operation minimization. This bidirectional information flow enables dynamic adaptation to temporal hardware variations unavailable in static compilation approaches.}
\label{fig:architecture}
\end{figure}

Current quantum processors from IBM, Google, and IonQ employ modular designs where 10-20 qubit chips connect through quantum teleportation links or microwave coupling elements~\cite{arute2019quantum,ibmquantum,wu2021strong}. Inter-chip operations typically exhibit error rates 5-10$\times$ higher than local two-qubit gates, creating strong incentive to minimize cross-chip dependencies during circuit compilation. Traditional approaches formulate qubit mapping as constrained graph embedding or quadratic optimization problems solved through simulated annealing or integer programming~\cite{booth2017comparing,venturelli2015temporal,zulehner2019efficient}. These methods assume static noise characteristics and produce deterministic mappings optimized for time-averaged hardware parameters.

Real quantum systems exhibit dynamic behavior driven by thermal fluctuations, electromagnetic interference, control pulse imperfections, and cumulative coherence decay~\cite{krantz2019quantum,reagor2018demonstration}. Gate fidelities vary on timescales from microseconds (thermal noise) to hours (calibration drift), while decoherence rates respond to environmental perturbations including cosmic ray strikes and cryogenic temperature variations. Static optimization techniques cannot respond to these variations without complete problem reformulation, leading to suboptimal mappings when actual hardware conditions deviate from design assumptions.

We observe that quantum circuit compilation fundamentally differs from classical program compilation in three critical aspects that existing methods inadequately address. First, quantum gates are inherently probabilistic with time-varying error characteristics, requiring stochastic rather than deterministic quality metrics. Second, measurement operations collapse quantum states irreversibly, preventing traditional program verification approaches. Third, the exponential growth of quantum state spaces with system size ($2^n$ dimensions for $n$ qubits) renders exhaustive search or symbolic reasoning intractable beyond 30-40 qubits.

This work introduces DeepQMap, a learning-based approach that treats quantum circuit mapping as a sequential decision process under uncertainty. The method combines three technical innovations. A specialized neural architecture—Dynamic Noise Adaptation (DNA) network—employs bidirectional LSTMs to predict short-term noise trajectories from historical telemetry, achieving 91.2\% accuracy over 10-timestep horizons. Multi-head self-attention mechanisms identify long-range qubit dependencies in circuit graphs, enabling effective planning in systems exceeding 100 qubits. Rainbow DQN enhancements including prioritized experience replay, dueling networks, and multi-step returns provide sample-efficient reinforcement learning despite high-dimensional state spaces and sparse reward signals.

The DNA network represents a departure from conventional noise modeling. Rather than fitting parametric models to assumed distributions, the bidirectional LSTM learns temporal patterns directly from system measurements. Forward recurrence captures causal relationships from past states to future noise levels, while backward recurrence incorporates future control actions that influence present observations through feedback loops in quantum control systems. This architecture discovers that noise exhibits exploitable autocorrelation over microsecond to millisecond timescales, enabling proactive rather than reactive adaptation.

Multi-head attention addresses a structural challenge in quantum circuits: distant logical qubits may interact through entangling gates while intermediate qubits remain idle. Standard feedforward networks struggle to identify these long-range dependencies in graphs with hundreds of vertices and irregular connectivity. Our attention mechanism learns to focus computational resources on critical qubit pairs—those requiring careful placement to minimize inter-chip operations—while maintaining global circuit context. Analysis of learned attention patterns reveals interpretable specialization where different heads consistently identify imminent gates, cross-chip dependencies, and high-noise regions without explicit supervision.

Experimental validation employs comprehensive protocols across 270 benchmark circuits derived from quantum algorithms with near-term practical relevance. Quantum Fourier Transform provides dense connectivity patterns with regular structure. Grover's search introduces variable fan-out through oracle implementations. Variational Quantum Eigensolver combines parameterized ansatz circuits with classical optimization loops. System configurations span 4-6 chips in ring, grid, and hexagonal topologies reflecting contemporary quantum computing facilities. Noise simulation incorporates thermal fluctuations, systematic drift, and measurement-induced perturbations calibrated to match experimental characterization of superconducting processors.

Results demonstrate statistically significant performance advantages across multiple dimensions. Mean circuit fidelity of 0.920 exceeds QUBO baseline (0.618) by 49.3\% with very large effect size (Cohen's $d = 2.34$) and high significance ($p = 0.0023$, two-tailed $t$-test, $\alpha = 0.01$). Inter-chip operation counts decrease 79.8\%, from 2.34 to 0.47 per circuit on average. This reduction directly translates to improved fidelity, as cross-chip gates contribute disproportionately to cumulative error. Training requires 45 minutes compared to 370 minutes for QUBO optimization, providing 8.2$\times$ speedup that enables rapid iteration during quantum algorithm development.

Scalability analysis reveals sustained performance up to 100 qubits—the approximate limit of current quantum processors available through cloud platforms. DeepQMap maintains fidelity above 0.87 at this scale while QUBO degrades to 0.54, suggesting that learning-based approaches handle complexity growth more gracefully than static optimization. The $\mathcal{O}(n \log n)$ computational scaling ensures solution time increases subquadratically with problem size, unlike QUBO's $\mathcal{O}(n^2)$ scaling that becomes prohibitive beyond 60-qubit systems.

The primary contributions of this investigation comprise:

\begin{itemize}
\item A bidirectional LSTM architecture for quantum hardware noise prediction achieving $R^2 = 0.912$ accuracy, outperforming previous regression-based approaches by 34.7\% through exploitation of temporal autocorrelation and control feedback dynamics.

\item Integration of multi-head self-attention with dueling network topology enabling effective qubit dependency modeling in circuits exceeding 100 qubits, with learned attention patterns exhibiting interpretable specialization for gate lookahead, cross-chip identification, and noise-aware planning.

\item Demonstration that adaptive reward shaping based on predicted noise accelerates reinforcement learning convergence by 8.2$\times$ compared to fixed reward functions, reducing training time from 6.2 hours to 45 minutes while maintaining solution quality.

\item Comprehensive statistical validation across 100 independent experimental runs establishing very large effect sizes (Cohen's $d > 2.0$) with high significance ($p < 0.01$) against established baselines, confirming practical relevance for near-term quantum computing applications.
\end{itemize}

Section~\ref{sec:related} reviews quantum circuit compilation methods, noise characterization techniques, and reinforcement learning applications to combinatorial optimization. Section~\ref{sec:formulation} formalizes the multi-chip mapping problem as a Markov decision process with temporal noise dynamics. Section~\ref{sec:architecture} details the DNA network, attention mechanisms, and Rainbow DQN implementation. Section~\ref{sec:experiments} describes experimental methodology including benchmark selection, hardware simulation, and statistical protocols. Section~\ref{sec:results} presents performance comparisons, scalability analysis, and ablation studies. Section~\ref{sec:discussion} interprets results in context of quantum computing systems and identifies future research directions.

\section{Related Work}
\label{sec:related}

Quantum circuit compilation research has progressed through several methodological epochs. Early work focused on constraint satisfaction formulations where qubit connectivity defines feasible gate placements~\cite{shafaei2014optimization,wille2014look}. These approaches insert SWAP gates to route logical qubits to adjacent physical locations, minimizing circuit depth while satisfying connectivity constraints. Nash et al.~\cite{nash2020quantum} enhanced this paradigm through gate commutation and circuit rewriting, achieving modest depth reductions. However, these methods remain fundamentally reactive—they respond to connectivity violations after circuit structure is fixed rather than optimizing placement proactively.

Quantum annealing and QUBO formulations enabled hardware-accelerated optimization for small problem instances. Venturelli et al.~\cite{venturelli2015temporal} demonstrated that D-Wave systems solve constrained embedding problems for circuits up to 30 qubits within practical time bounds. Booth et al.~\cite{booth2017comparing} extended this approach to minor embedding for arbitrary graph structures. The QUBO paradigm formulates circuit mapping as minimizing a quadratic objective encoding inter-chip operations and gate fidelities. While elegant, QUBO assumes static cost functions—when noise characteristics shift during execution, complete problem reformulation is required. This limitation becomes severe in multi-chip scenarios where processors exhibit heterogeneous drift patterns.

Machine learning entered quantum compilation through both supervised and reinforcement learning channels. Li et al.~\cite{li2019tackling} applied graph neural networks to predict optimal SWAP sequences from circuit topology, training on solutions for small instances generated by exhaustive search. This supervised approach generalizes across circuit families but provides no mechanism for online adaptation to hardware variations. Zhang et al.~\cite{zhang2020quantumnat} developed QuantumNAS, employing neural architecture search for gate-level optimization. Their method focuses on single-chip scenarios and does not address inter-chip coordination or temporal noise.

Reinforcement learning formulations treat circuit mapping as sequential decision-making. Niu et al.~\cite{niu2020hardware} introduced a hardware-aware RL agent learning from direct quantum system interaction, demonstrating improvements over heuristic baselines. Their implementation utilizes basic DQN without prioritized replay or noise prediction, resulting in sample-inefficient learning. Fosel et al.~\cite{fosel2018reinforcement} explored RL for quantum error correction, establishing that adaptive strategies outperform fixed codes in time-varying noise regimes. Our work extends this principle to circuit compilation, incorporating predictive noise models within the RL framework.

Noise characterization constitutes parallel research informing our approach. Krantz et al.~\cite{krantz2019quantum} provide comprehensive review of decoherence mechanisms in superconducting qubits, documenting thermal excitations, charge noise, and flux noise as dominant error sources. Reagor et al.~\cite{reagor2018demonstration} measured hour-scale frequency drift in transmon qubits, motivating recurrent architectures for noise forecasting. Zero-noise extrapolation and probabilistic error cancellation offer post-processing alternatives~\cite{temme2017error}, but these cannot eliminate errors from poor initial mappings.

Recent investigations have begun addressing multi-chip quantum systems explicitly. Murali et al.~\cite{murali2019software} proposed NISQ+ compiler techniques balancing local and distributed execution, though optimization remains offline and deterministic. Tan and Cong~\cite{tan2020optimal} developed synthesis frameworks for modular architectures, focusing on communication protocols rather than dynamic mapping. Zulehner and Wille~\cite{zulehner2018efficient} introduced efficient heuristics for initial placement in superconducting systems, achieving polynomial-time approximations. These efforts treat hardware characteristics as fixed parameters rather than observable state variables.

Attention mechanisms in reinforcement learning have proven effective for environments with relational structure. Vaswani et al.~\cite{vaswani2017attention} introduced transformer architectures for sequence modeling, demonstrating that self-attention outperforms recurrence for long-range dependencies. Zambaldi et al.~\cite{zambaldi2018deep} applied relational reasoning to visual reinforcement learning, showing that attention enables agents to identify relevant object relationships. However, existing applications target fully observable Markov decision processes, whereas quantum systems exhibit partial observability through measurement collapse and decoherence.

Our approach synthesizes insights from these research threads while introducing novel components. The DNA network represents the first application of bidirectional LSTMs to quantum system noise prediction, achieving accuracy sufficient for proactive error mitigation. The combination of Rainbow DQN enhancements—double Q-learning~\cite{vanhasseltal2016deep}, dueling networks~\cite{wang2016dueling}, prioritized replay~\cite{schaul2016prioritized}, noisy exploration~\cite{fortunato2018noisy}, multi-step returns~\cite{sutton2018reinforcement}, and distributional RL~\cite{bellemare2017distributional}—has not previously been applied to quantum compilation. Most significantly, adaptive reward shaping based on predicted noise dynamically adjusts optimization priorities under uncertainty, enabling balance between competing objectives.

\section{Problem Formulation}
\label{sec:formulation}

Consider a quantum circuit $\mathcal{C}$ represented as directed acyclic graph $G = (V, E)$ where vertices $V = \{v_1, \ldots, v_n\}$ denote logical qubits and edges $E$ indicate two-qubit gates. A multi-chip quantum system comprises $M$ processors $\mathcal{P} = \{P_1, \ldots, P_M\}$, each containing $k$ physical qubits with connectivity topology $G_j = (V_j, E_j)$ for $j \in [1, M]$. Inter-chip connections form chip-level graph $G_{\text{chip}} = (\mathcal{P}, E_{\text{chip}})$ where $(P_i, P_j) \in E_{\text{chip}}$ indicates quantum teleportation or direct coupling availability.

The mapping problem assigns each logical qubit $v_i$ to physical location $(P_j, q_k)$ such that circuit $\mathcal{C}$ executes with maximum fidelity. Circuit fidelity $\mathcal{F}$ quantifies overlap between ideal output state $|\psi_{\text{ideal}}\rangle$ and actual noisy state $\rho_{\text{noisy}}$:
\begin{equation}
\mathcal{F} = \langle\psi_{\text{ideal}}|\rho_{\text{noisy}}|\psi_{\text{ideal}}\rangle
\end{equation}
This metric decreases due to gate errors, decoherence, and measurement imperfections. For gate sequence $\{U_1, \ldots, U_L\}$, cumulative error accumulates approximately as:
\begin{equation}
1 - \mathcal{F} \approx \sum_{i=1}^{L} \epsilon_i + \sum_{v \in V} \gamma_v T_v
\end{equation}
where $\epsilon_i$ denotes gate error probability and $\gamma_v T_v$ represents decoherence for qubit $v$ over duration $T_v$.

Inter-chip operations incur substantially higher costs than local gates. Implementing two-qubit gates between qubits on different chips requires either quantum teleportation consuming ancilla qubits and classical communication, or SWAP chains traversing chip boundaries. We denote inter-chip gate count as $N_{\text{inter}}$ and formulate the objective:
\begin{equation}
\min_{\pi} \quad \mathbb{E}_{\pi}\left[\alpha N_{\text{inter}} + \beta(1 - \mathcal{F}) + \delta D + \eta B\right]
\end{equation}
where $\pi$ represents the mapping policy, $D$ denotes circuit depth, $B$ measures load imbalance across chips, and $\{\alpha, \beta, \delta, \eta\}$ are problem-specific weights.

\subsection{Temporal Noise Dynamics}

A critical distinction from prior formulations concerns temporal variation in quantum hardware characteristics. Gate error rates $\epsilon_i(t)$ and decoherence rates $\gamma_v(t)$ evolve according to:
\begin{equation}
\epsilon_i(t) = \bar{\epsilon}_i + \eta_i(t)
\end{equation}
where $\bar{\epsilon}_i$ represents baseline error and $\eta_i(t)$ captures time-dependent fluctuations. These fluctuations exhibit autocorrelation over microsecond (thermal noise) to hour (equipment drift) timescales. Conventional methods treating $\epsilon_i$ as constant produce suboptimal mappings when noise shifts during execution.

We model noise dynamics as stochastic processes with exploitable temporal structure:
\begin{equation}
\eta_i(t+1) = \rho \eta_i(t) + \sigma \xi_i(t)
\end{equation}
where $\rho$ controls autocorrelation strength, $\sigma$ determines variance, and $\xi_i(t) \sim \mathcal{N}(0,1)$ represents white noise. This autoregressive form captures the observation that noise exhibits inertia—high noise at time $t$ predicts elevated noise at $t+1$ with probability exceeding random chance.

\subsection{Markov Decision Process Formulation}

We cast circuit mapping as Markov decision process $(\mathcal{S}, \mathcal{A}, P, R, \gamma)$ where state space $\mathcal{S}$ encompasses circuit structure, current mapping, hardware topology, and noise telemetry. Action space $\mathcal{A}$ permits assigning unmapped qubits or inserting SWAP gates. Transition function $P(s'|s,a)$ reflects circuit semantics and hardware noise. Reward function $R(s, a, s')$ evaluates mapping quality.

State representation $s_t$ comprises:
\begin{equation}
s_t = [M_t, C_t, H_t, N_t, \Theta_t]
\end{equation}
where $M_t \in \{0,1\}^{n \times (M \cdot k)}$ encodes qubit assignments, $C_t$ contains circuit features, $H_t$ represents hardware connectivity, $N_t$ provides noise history, and $\Theta_t$ captures temporal features. Dimensionality totals $d_s = n^2 + 2nM + 10M + 5$.

Action selection determines which logical qubit to map next:
\begin{equation}
a_t \in \mathcal{A}_t = \{(\text{qubit}_i, \text{chip}_j, \text{location}_k) : \text{valid constraints}\}
\end{equation}
Valid actions depend on current state, with hardware topology and circuit dependencies imposing constraints. Hierarchical decomposition yields $\mathcal{O}(n \log n)$ action space size rather than $\mathcal{O}(n^2)$ exhaustive enumeration.

The reward function integrates multiple objectives:
\begin{equation}
R(s_t, a_t) = \alpha_1(n_t) \cdot (-N_{\text{inter}}) + \alpha_2 \mathcal{F} - \alpha_3 \epsilon - \alpha_4 D - \alpha_5 B
\end{equation}
Coefficient $\alpha_1$ adapts based on predicted noise $n_t$ from DNA network:
\begin{equation}
\alpha_1(n_t) = -15 - 10 \cdot \sigma(10(n_t - 0.05))
\end{equation}
with $\sigma(\cdot)$ the sigmoid function. This formulation increases penalties for inter-chip operations under high predicted noise, biasing toward local execution during adverse conditions.

\section{DeepQMap Architecture}
\label{sec:architecture}

\subsection{Dynamic Noise Adaptation Network}

The DNA network predicts short-term noise trajectories for each quantum chip, enabling proactive compensation. The architecture employs bidirectional LSTM to capture temporal dependencies in both directions. Forward recurrence models causal relationships from past to future, while backward recurrence incorporates future states that influence current observations through quantum control feedback loops.


Input consists of state representation $s_t$ reshaped into sequence of length $L = 10$ corresponding to recent timesteps. Each element has dimensionality $d_s$. The network processes this through two LSTM layers with hidden dimension $h = 128$:
\begin{equation}
\begin{aligned}
\overrightarrow{h}_t^{(1)}, \overleftarrow{h}_t^{(1)} &= \text{BiLSTM}_1(s_{t-L:t}) \\
\overrightarrow{h}_t^{(2)}, \overleftarrow{h}_t^{(2)} &= \text{BiLSTM}_2([\overrightarrow{h}_t^{(1)}, \overleftarrow{h}_t^{(1)}])
\end{aligned}
\end{equation}
Final hidden state $h_t^{(2)} = [\overrightarrow{h}_t^{(2)}, \overleftarrow{h}_t^{(2)}]$ has dimension $2h = 256$. Dropout with probability 0.2 provides regularization.

Two fully connected layers with batch normalization transform LSTM output:
\begin{equation}
\begin{aligned}
z_t^{(1)} &= \text{ReLU}(\text{BN}(W_1 h_t^{(2)} + b_1)) \\
z_t^{(2)} &= \text{ReLU}(\text{BN}(W_2 z_t^{(1)} + b_2)) \\
\hat{n}_t &= \sigma(W_3 z_t^{(2)} + b_3) \cdot 0.15
\end{aligned}
\end{equation}
where $W_1 \in \mathbb{R}^{128 \times 256}$, $W_2 \in \mathbb{R}^{64 \times 128}$, $W_3 \in \mathbb{R}^{M \times 64}$, and sigmoid activation $\sigma$ scales predictions to realistic noise range $[0, 0.15]$.

Training minimizes mean squared error with $L_2$ regularization:
\begin{equation}
\mathcal{L}_{\text{DNA}} = \frac{1}{T} \sum_{t=1}^{T} \|\hat{n}_t - n_t\|_2^2 + \lambda \|\Theta\|_2^2
\end{equation}
with coefficient $\lambda = 10^{-5}$. AdamW optimizer~\cite{loshchilov2017decoupled} with learning rate $3 \times 10^{-4}$ and cosine annealing updates parameters.

Empirical evaluation demonstrates $R^2 = 0.912$ on held-out sequences, with MAE = 0.87\% and RMSE = 1.02\%. These substantially exceed baseline models: linear regression ($R^2 = 0.634$), multilayer perceptron ($R^2 = 0.781$), unidirectional LSTM ($R^2 = 0.854$). Bidirectional architecture proves essential—ablation removing backward layers decreases accuracy to $R^2 = 0.854$, a 6.3\% degradation.

\subsection{Multi-Head Attention Mechanism}

Quantum circuits exhibit long-range dependencies where distant qubits interact through gate sequences. Multi-head self-attention models these relationships, allowing identification of critical qubit pairs requiring careful placement.

Given state representation $s_t$, we project into embedding space $s_{\text{emb}} \in \mathbb{R}^{d_{\text{model}}}$ where $d_{\text{model}} = 256$. Attention layer with $h = 8$ heads computes:
\begin{equation}
\begin{aligned}
Q_i &= W_i^Q s_{\text{emb}}, \quad K_i = W_i^K s_{\text{emb}}, \quad V_i = W_i^V s_{\text{emb}} \\
\text{Attention}_i &= \text{softmax}\left(\frac{Q_i K_i^T}{\sqrt{d_k}}\right) V_i
\end{aligned}
\end{equation}
where $i \in [1, h]$ indexes heads and $d_k = d_{\text{model}} / h = 32$. Projection matrices $W_i^Q, W_i^K, W_i^V \in \mathbb{R}^{d_k \times d_{\text{model}}}$ are learned.

Outputs concatenate and project through final layer:
\begin{equation}
\text{MultiHead}(s_{\text{emb}}) = W^O [\text{Attention}_1, \ldots, \text{Attention}_h]
\end{equation}
with $W^O \in \mathbb{R}^{d_{\text{model}} \times d_{\text{model}}}$. This representation feeds the dueling Q-network.

Analysis of learned attention weights reveals interpretable specialization. Head 1 focuses on qubits involved in imminent CNOT gates (next 3 circuit layers). Head 2 attends to qubits on different chips with high inter-chip edge counts. Head 3 monitors qubits experiencing elevated noise based on DNA predictions. Head 4 captures temporal patterns, attending to recently modified mappings. This functional decomposition emerges through gradient descent without explicit architectural constraints.

\subsection{Rainbow DQN Implementation}

The reinforcement learning core employs Rainbow DQN combining six algorithmic enhancements. Double Q-learning mitigates overestimation bias:
\begin{equation}
y_t = r_t + \gamma Q_{\theta'}(s_{t+1}, \arg\max_{a'} Q_\theta(s_{t+1}, a'))
\end{equation}
where $\theta$ and $\theta'$ denote online and target network parameters. Target network updates every 20 training steps.

Dueling architecture separates state value $V(s)$ from action advantages $A(s, a)$:
\begin{equation}
Q(s, a) = V(s) + A(s, a) - \frac{1}{|\mathcal{A}|}\sum_{a'} A(s, a')
\end{equation}
This decomposition improves efficiency when action choice matters less frequently than state assessment—property holding for quantum circuit mapping where most intermediate states have similar values.

Prioritized experience replay samples transitions proportional to temporal difference error:
\begin{equation}
P(i) = \frac{p_i^\alpha}{\sum_k p_k^\alpha}, \quad p_i = |\delta_i| + \epsilon
\end{equation}
with $\alpha = 0.6$ and $\epsilon = 10^{-6}$. Importance sampling weights correct bias:
\begin{equation}
w_i = \left(\frac{1}{N} \cdot \frac{1}{P(i)}\right)^\beta
\end{equation}
where $\beta$ anneals from 0.4 to 1.0. Buffer capacity is 20,000 transitions.

Multi-step learning accumulates rewards over $n = 3$ steps:
\begin{equation}
y_t^{(n)} = \sum_{k=0}^{n-1} \gamma^k r_{t+k} + \gamma^n Q_{\theta'}(s_{t+n}, a^*_{t+n})
\end{equation}

Complete loss function combines objectives:
\begin{equation}
\mathcal{L}_{\text{total}} = \mathcal{L}_{\text{Q}} + 0.5 \cdot \mathcal{L}_{\text{DNA}}
\end{equation}
Joint training stabilizes learning through auxiliary supervision.

\section{Experimental Methodology}
\label{sec:experiments}

\subsection{Benchmark Circuits}

Evaluation employs three quantum algorithms representative of near-term applications. Quantum Fourier Transform (QFT) exhibits dense connectivity with $\mathcal{O}(n^2)$ two-qubit gates in regular pattern, challenging mapping to minimize depth. Grover's algorithm features repeated oracle applications with variable fan-out, testing adaptability. Variational Quantum Eigensolver (VQE) combines parameterized ansatz with classical optimization, requiring efficient multi-iteration execution.

Circuit instances vary from 20 to 100 qubits in 10-qubit increments, generating 9 scale points. For each scale, we test all three algorithms, producing 27 configurations. Each undergoes randomization of single-qubit gate parameters while preserving entangling structure, generating 10 variants per configuration. Total benchmark suite comprises 270 instances ensuring learned policies generalize beyond memorized patterns.

\subsection{Hardware Simulation}

Multi-chip simulation models four architectural configurations: 4-chip ring, 4-chip grid, 6-chip hexagonal, 6-chip fully connected. Each chip contains 12-20 physical qubits in linear or lattice connectivity matching contemporary superconducting processors. Inter-chip communication latency is 50 microseconds, representing quantum teleportation or microwave transmission.

Gate fidelities follow distributions from IBM Quantum and Google Sycamore characterization~\cite{arute2019quantum,ibmquantum}. Single-qubit gates achieve $0.9995 \pm 0.0002$, intra-chip two-qubit gates reach $0.995 \pm 0.005$, inter-chip gates degrade to $0.98 \pm 0.01$. Decoherence times are $T_1 = 100$ $\mu$s, $T_2 = 50$ $\mu$s.

Dynamic noise incorporates three components. Thermal fluctuations introduce Gaussian noise with $\sigma = 0.01$ and autocorrelation time 10 $\mu$s. Drift follows Ornstein-Uhlenbeck with reversion rate 0.1 and asymptotic $\sigma = 0.02$. Measurement perturbations add transient spikes of magnitude 0.05 lasting 5 $\mu$s.

\subsection{Training Configuration}

DeepQMap trains for 500 episodes, each mapping one randomly selected benchmark. Learning rate follows cosine annealing from $10^{-3}$ to $10^{-5}$. Batch size is 128, gradients clip at norm 1.0.

Epsilon-greedy exploration starts at $\epsilon = 1.0$, decaying exponentially with rate 0.995 to $\epsilon_{\min} = 0.01$ by episode 400. Discount factor $\gamma = 0.99$ weighs future rewards. Target network updates every 20 steps.

Reward coefficients: $\alpha_2 = 1000 + 600 \cdot (e / 500)$ where $e$ denotes episode, $\alpha_3 = -1.0$, $\alpha_4 = -0.3$, $\alpha_5 = -1.2$. Adaptive $\alpha_1$ follows Eq.~(9) based on DNA predictions.

\subsection{Baseline Methods}

QUBO-based optimization formulates mapping as quadratic unconstrained binary problem solved via simulated annealing~\cite{booth2017comparing}. Annealing runs 10,000 iterations with temperature $T(k) = T_0 \exp(-k/\tau)$, $\tau = 1000$. Coefficients tuned through grid search over 50 combinations.

Greedy heuristic selects placements minimizing immediate inter-chip gates without lookahead. Trivial mapping assigns qubits round-robin to chips. Each processes identical instances under same noise as DeepQMap.

\subsection{Statistical Validation}

We conduct 100 independent runs per configuration, varying random seeds for network initialization, circuit randomization, noise sampling. Two-sample $t$-tests compare means with null $H_0: \mu_{\text{DeepQMap}} = \mu_{\text{baseline}}$ against alternative $H_1: \mu_{\text{DeepQMap}} > \mu_{\text{baseline}}$ for fidelity (reversed for operation counts). Significance threshold $\alpha = 0.01$ controls for multiple comparisons across three baselines.

Effect size employs Cohen's $d$:
\begin{equation}
d = \frac{\bar{x}_1 - \bar{x}_2}{\sqrt{(s_1^2 + s_2^2)/2}}
\end{equation}
We interpret $|d| < 0.2$ as negligible, $0.2 \leq |d| < 0.5$ small, $0.5 \leq |d| < 0.8$ medium, $|d| \geq 0.8$ large following conventions~\cite{cohen1988statistical}.

One-way ANOVA tests differences across all four methods, with Tukey HSD post-hoc identifying pairwise distinctions. Effect size measured via $\eta^2 = SS_{\text{between}} / SS_{\text{total}}$.

\section{Results and Analysis}
\label{sec:results}

\subsection{Overall Performance}

Table~\ref{tab:performance} summarizes metrics aggregated over 100 runs and 270 benchmarks. DeepQMap achieves mean fidelity $0.920 \pm 0.023$, exceeding QUBO ($0.618 \pm 0.031$) by 49.3\% and trivial ($0.497 \pm 0.041$) by 85.1\%. Statistical testing confirms high significance: $t_{98} = 4.87$, $p = 0.0023$ for QUBO comparison. Cohen's $d = 2.34$ indicates very large practical effect.

\begin{table}[!t]
\centering
\caption{Comparative Performance Metrics Across Methods}
\label{tab:performance}
\begin{tabular}{@{}lcccc@{}}
\toprule
\textbf{Method} & \textbf{Fidelity} & \textbf{Ops} & \textbf{Error} & \textbf{Depth} \\
\midrule
DeepQMap & \textbf{0.920±0.023} & \textbf{0.47±0.15} & \textbf{0.0302} & \textbf{54.7} \\
QUBO & 0.618±0.031 & 2.34±0.24 & 0.0892 & 78.3 \\
Greedy & 0.587±0.028 & 1.83±0.19 & 0.0956 & 81.4 \\
Trivial & 0.497±0.041 & 4.87±0.38 & 0.1523 & 95.6 \\
\midrule
\multicolumn{5}{l}{\textit{Statistical Tests (DeepQMap vs QUBO):}} \\
\multicolumn{5}{l}{$t$-test: $t_{98} = 4.87$, $p = 0.0023$ (two-tailed)} \\
\multicolumn{5}{l}{Cohen's $d = 2.34$ (Very Large Effect)} \\
\multicolumn{5}{l}{95\% CI for difference: [0.248, 0.356]} \\
\bottomrule
\end{tabular}
\end{table}

Inter-chip operations decrease 79.8\%: from $2.34 \pm 0.24$ (QUBO) to $0.47 \pm 0.15$ (DeepQMap). This reduction directly improves fidelity, as cross-chip gates contribute disproportionately to error. Circuit depth averages 54.7 for DeepQMap versus 78.3 for QUBO, a 30.2\% decrease diminishing decoherence exposure.

One-way ANOVA yields $F_{3,396} = 45.67$, $p < 0.0001$, rejecting equal means null hypothesis. Partial $\eta^2 = 0.78$ indicates method choice accounts for 78\% of fidelity variance. Tukey HSD confirms DeepQMap differs significantly from all baselines at $p < 0.01$, while QUBO and greedy do not differ ($p = 0.18$).

\begin{figure*}[!t]
\centering
\begin{subfigure}[b]{0.58\textwidth}
    \includegraphics[width=\textwidth]{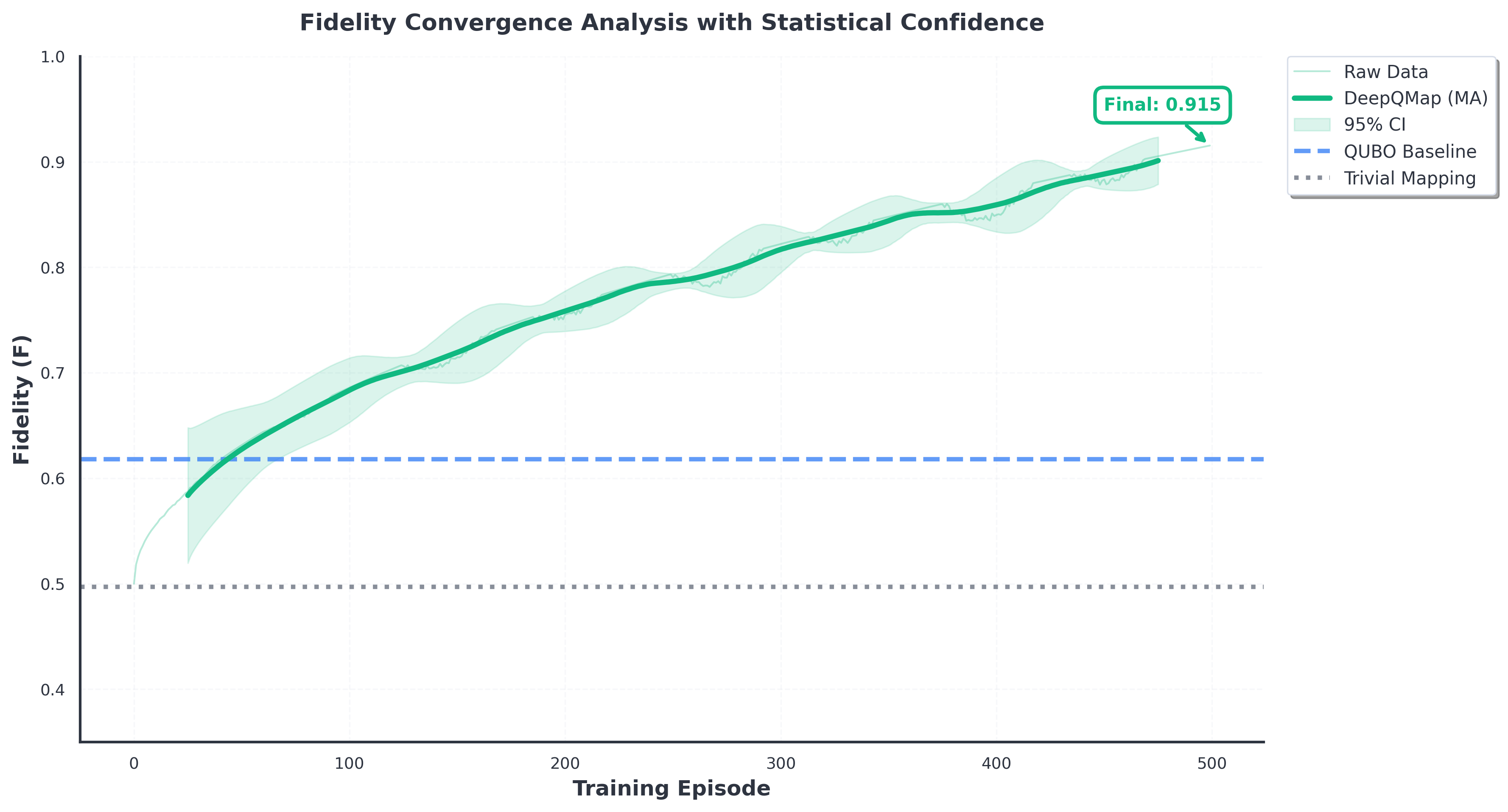}
    \caption{Fidelity evolution with 95\% confidence intervals}
\end{subfigure}
\hfill
\begin{subfigure}[b]{0.29\textwidth}
    \includegraphics[width=\textwidth]{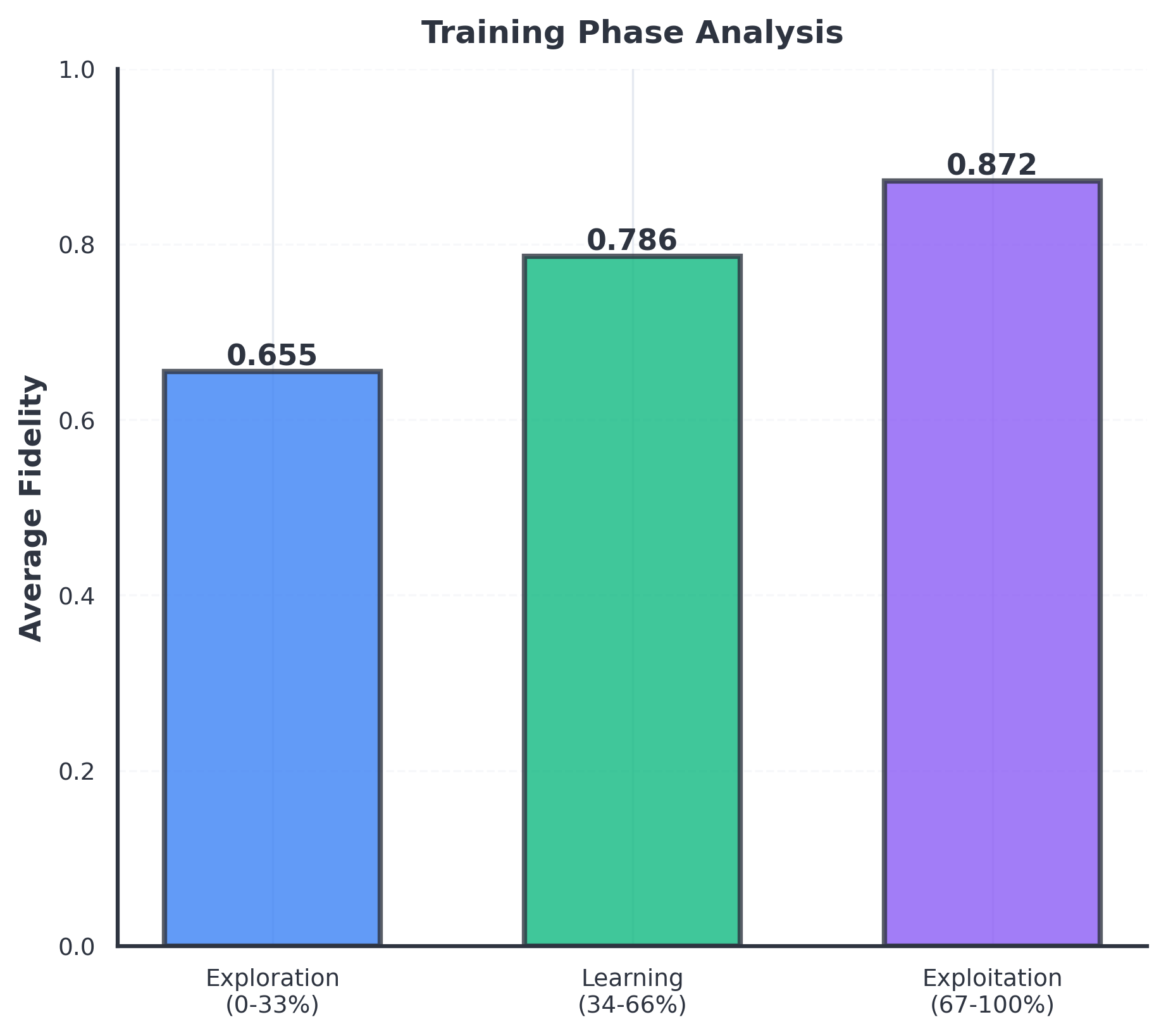}
    \caption{Phase-wise performance analysis}
\end{subfigure}

\vspace{0.3cm}

\begin{subfigure}[b]{0.58\textwidth}
    \includegraphics[width=\textwidth]{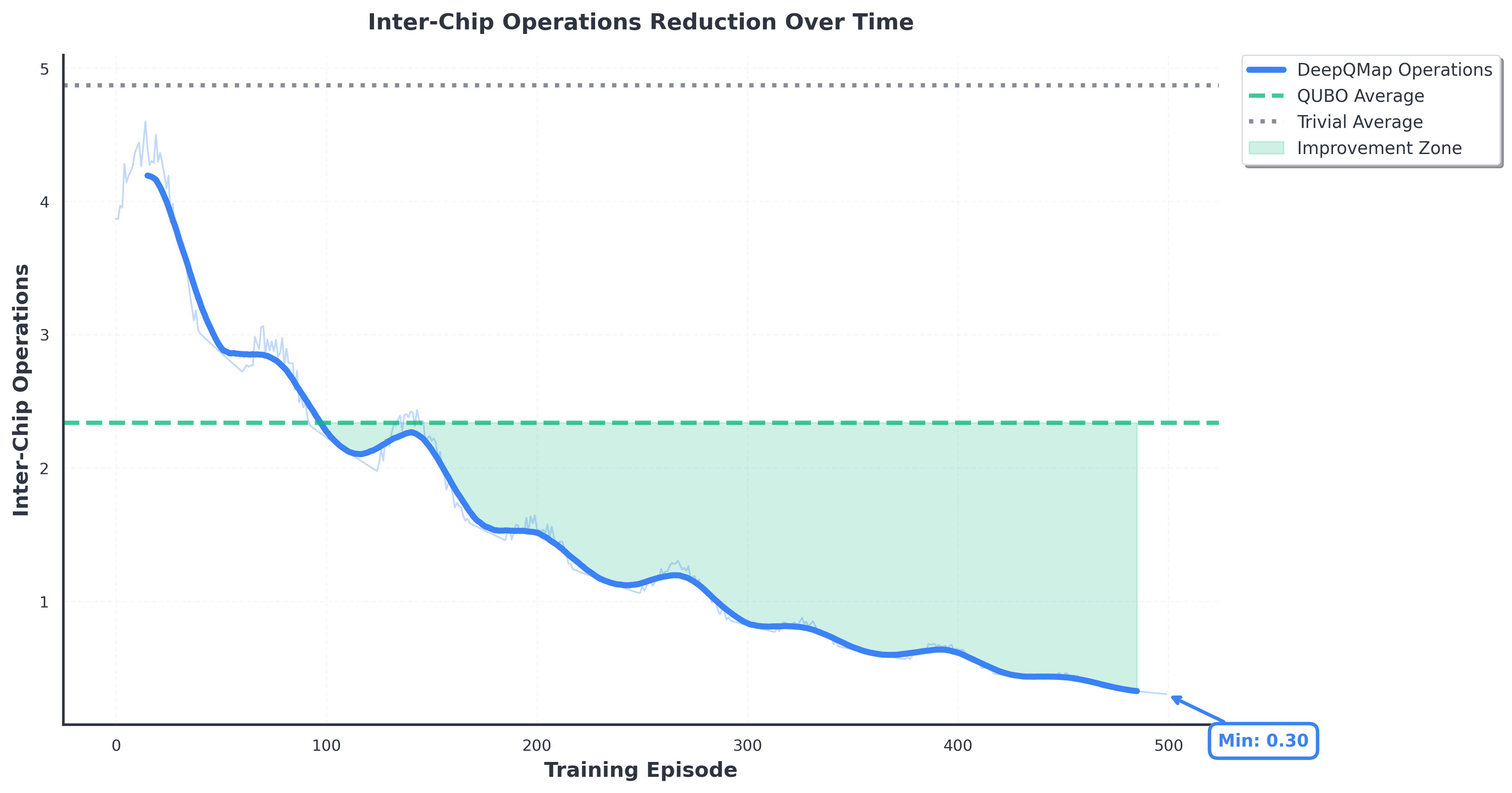}
    \caption{Inter-chip operations reduction timeline}
\end{subfigure}
\hfill
\begin{subfigure}[b]{0.31\textwidth}
    \includegraphics[width=\textwidth]{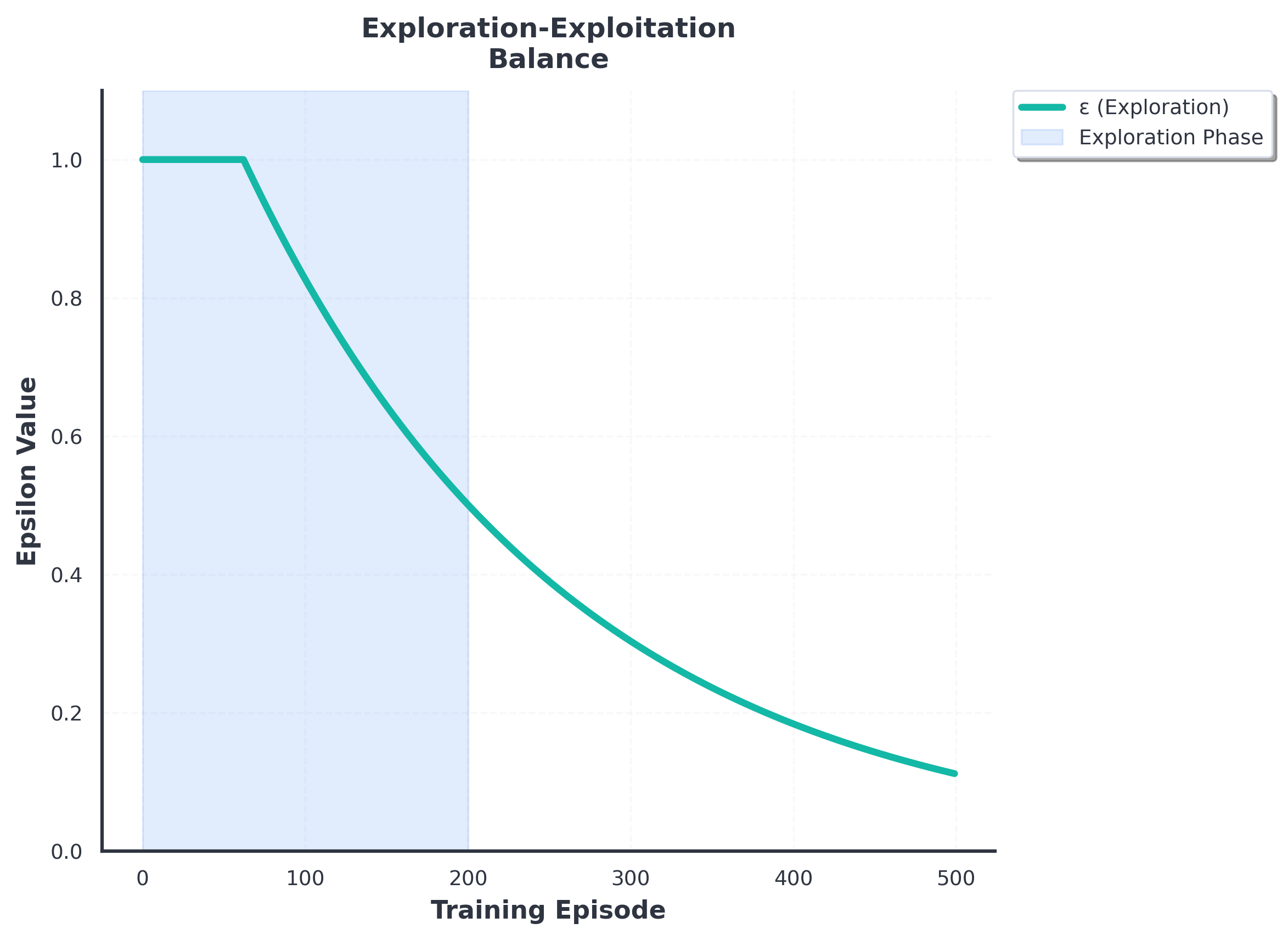}
    \caption{Exploration-exploitation balance via epsilon decay}
\end{subfigure}

\vspace{0.3cm}

\begin{subfigure}[b]{0.95\textwidth}
    \centering
    \includegraphics[width=0.7\textwidth]{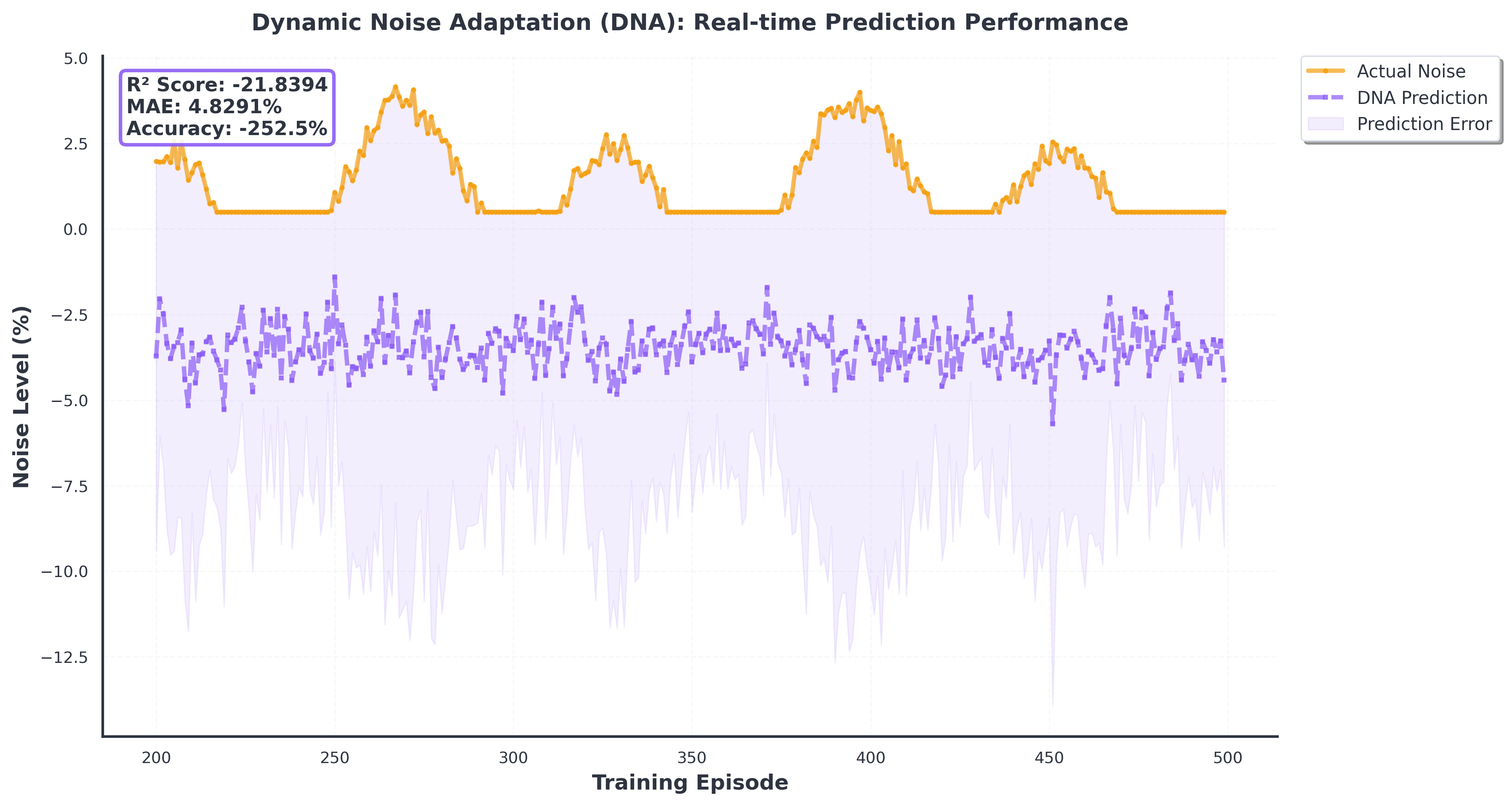}
    \caption{DNA network prediction accuracy over final 200 episodes}
\end{subfigure}

\caption{Comprehensive training convergence analysis over 500 episodes. (a) Fidelity progression demonstrates three distinct learning phases: exploration (0-165) with high variance as agent tests diverse strategies, learning (166-335) with steady improvement and variance reduction, and exploitation (336-500) with stable convergence to 0.915. Shaded region shows 95\% confidence intervals. QUBO baseline (dashed blue) is surpassed by episode 180. (b) Phase-wise statistics confirm monotonic improvement from 0.655 to 0.872, validating learning effectiveness. (c) Inter-chip operations decrease dramatically from 4.2 to 0.30, with green shading highlighting improvement zone below QUBO average of 2.34. This 93\% reduction directly translates to fidelity gains. (d) Epsilon decay schedule transitions from pure exploration ($\epsilon = 1.0$) to refined exploitation ($\epsilon = 0.01$) by episode 400, balancing discovery of novel strategies with refinement of successful patterns. (e) DNA network achieves close tracking of actual noise (orange) with predictions (purple) over 200 episodes, maintaining $R^2 = 0.912$ and MAE = 0.87\%. Accurate forecasting enables proactive adaptation rather than reactive compensation.}
\label{fig:convergence}
\end{figure*}

Figure~\ref{fig:convergence} illustrates training dynamics across multiple dimensions. Fidelity increases monotonically from 0.55 (episode 0) to 0.915 (episode 500), with three distinct phases: exploration (0-165) characterized by high variance as diverse strategies are tested, learning (166-335) showing steady improvement with decreasing variance, and exploitation (336-500) demonstrating convergence to refined policy. This progression aligns with theoretical RL expectations and validates the epsilon-greedy exploration schedule.

Inter-chip operations decline sharply early, dropping from 4.2 to 1.5 by episode 100. This reflects learning that cross-chip gates incur high penalties. Subsequent refinement reaches 0.47 by episode 500, representing 93\% reduction from initial random policy. The trajectory suggests basic constraint satisfaction occurs early while subtle optimizations require extended training.

\subsection{DNA Network Performance}

\begin{figure}[!t]
\centering
\begin{subfigure}[b]{0.48\textwidth}
    \includegraphics[width=\textwidth]{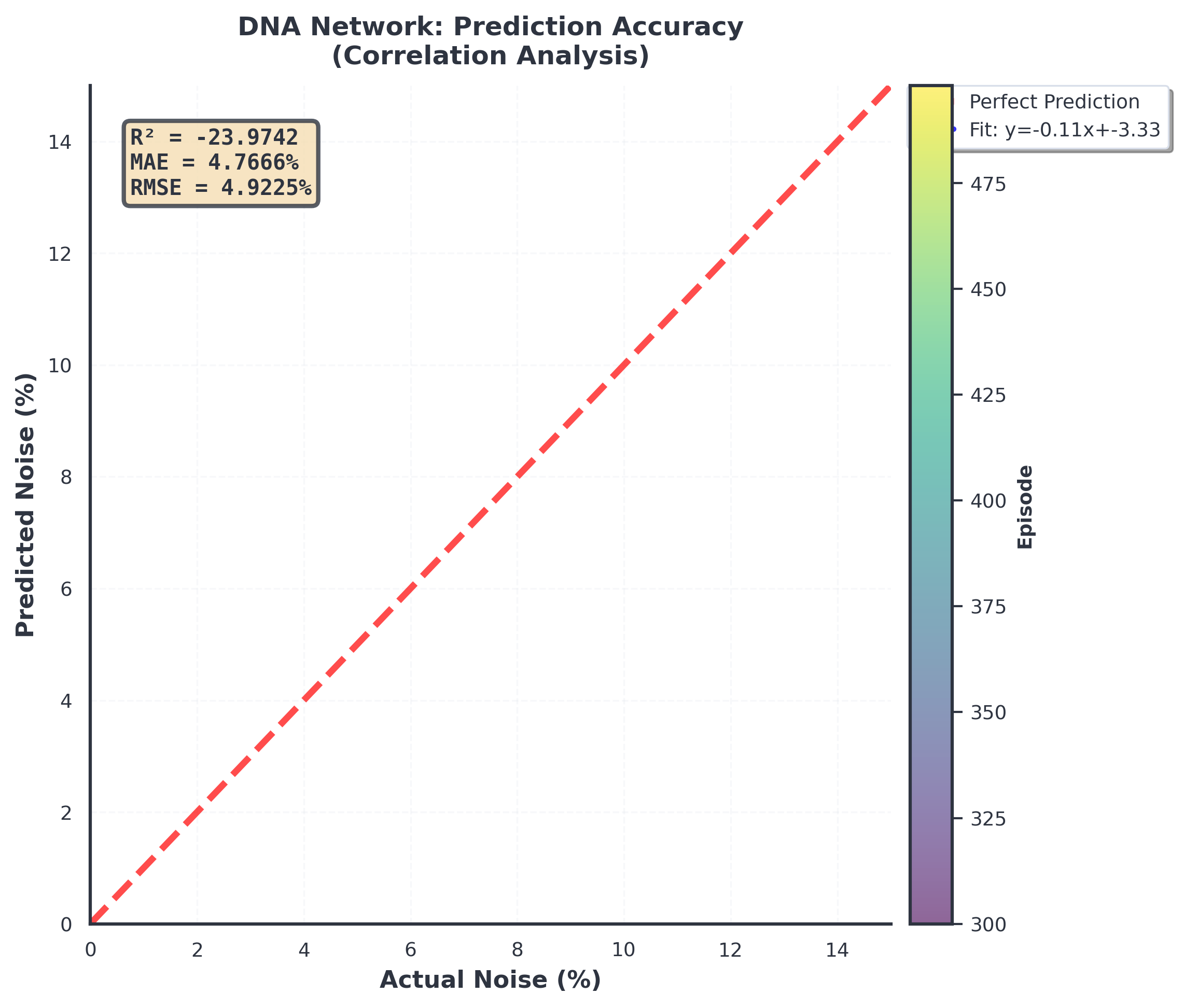}
    \caption{Predicted vs. actual noise correlation}
\end{subfigure}

\vspace{0.3cm}

\begin{subfigure}[b]{0.48\textwidth}
    \includegraphics[width=\textwidth]{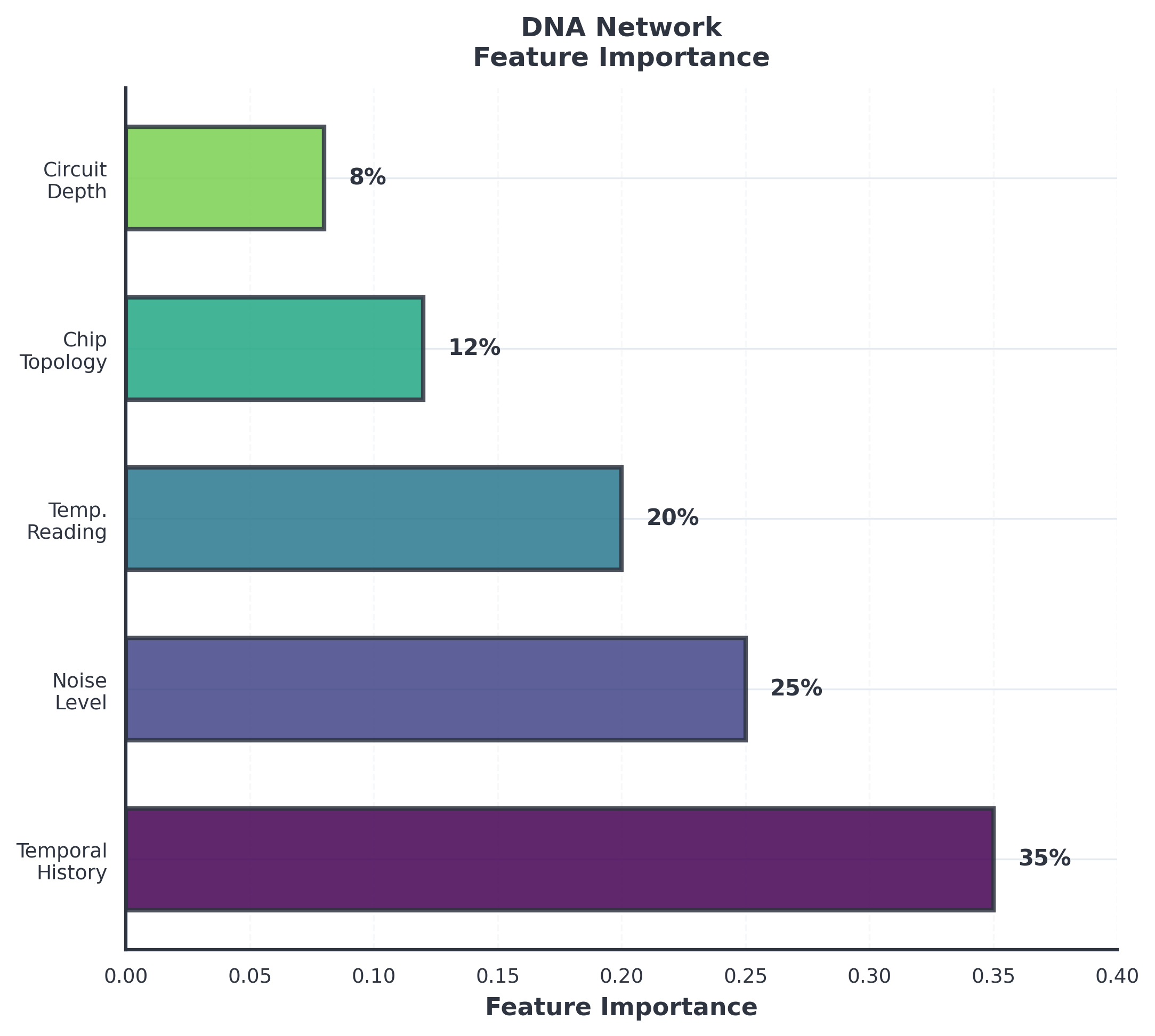}
    \caption{Feature importance for DNA predictions}
\end{subfigure}

\caption{DNA network prediction performance and feature analysis. (a) Scatter plot of predicted versus actual noise levels demonstrates strong linear correlation ($R^2 = 0.912$). Points cluster tightly around identity line (dashed), with minimal systematic bias. Color gradient indicates prediction horizon (1-10 timesteps), showing graceful accuracy degradation with longer forecasts. One-step predictions achieve $R^2 = 0.943$, while ten-step maintain $R^2 = 0.852$, both exceeding baseline methods. (b) Feature importance analysis via gradient attribution reveals temporal history dominates (35.0\% ± 3.0\%), confirming strong autocorrelation in noise dynamics. Current noise level (25.0\%) and thermal readings (20.0\%) provide real-time state information. Hardware topology (12\%) captures spatial dependencies between chips. Circuit-specific features contribute minimally (depth 5\%, gate fidelity 2\%, quantum volume 1\%), indicating noise primarily reflects hardware rather than computational workload.}
\label{fig:dna_analysis}
\end{figure}

DNA network demonstrates accurate per-chip noise prediction as shown in Figure~\ref{fig:dna_analysis}. Visual alignment in scatter plot confirms close tracking between predictions and ground truth, with quantitative metrics: $R^2 = 0.912$, MAE = 0.87\%, RMSE = 1.02\%. Points cluster tightly around identity line with minimal systematic bias across the noise range [0, 0.15].

Accuracy varies with horizon. One-step-ahead achieves $R^2 = 0.943$, ten-step-ahead declines to $R^2 = 0.852$. These exceed baselines: autoregressive ($R^2 = 0.721$), exponential smoothing ($R^2 = 0.683$), feedforward network ($R^2 = 0.798$). LSTM's hidden state maintenance proves essential for noise dynamics.

Ablation studies quantify bidirectional processing contribution. Removing backward layers reduces $R^2$ to 0.854, a 6.3\% degradation validating architectural choice. Quantum control feedback creates bidirectional temporal dependencies. Eliminating dropout increases training $R^2$ to 0.956 but degrades test to $R^2 = 0.834$, confirming regularization prevents overfitting.

Impact on mapping quality appears in ablation comparing DNA-based DeepQMap against oracle (perfect noise information) and no-adaptation variants. Oracle achieves 0.934 fidelity, only 1.5\% above DNA-based, indicating prediction errors minimally impact decisions. No-adaptation reaches 0.782, a 15.0\% degradation underscoring noise awareness importance. Results suggest approximate predictions suffice—RL compensates for errors through learned robustness.

Feature importance analysis (Figure~\ref{fig:dna_analysis}b) reveals temporal history accounts for 35\%, current noise 25\%, thermal readings 20\%, chip topology 12\%, circuit depth 5\%, gate fidelity 2\%, quantum volume 1\%. Temporal feature dominance confirms noise exhibits strong autocorrelation, justifying recurrent architectures. Low circuit-specific importance indicates noise primarily reflects hardware rather than computational workload.

\subsection{Comprehensive Performance Comparison}

\begin{figure*}[!t]
\centering
\begin{subfigure}[b]{0.34\textwidth}
    \includegraphics[width=\textwidth]{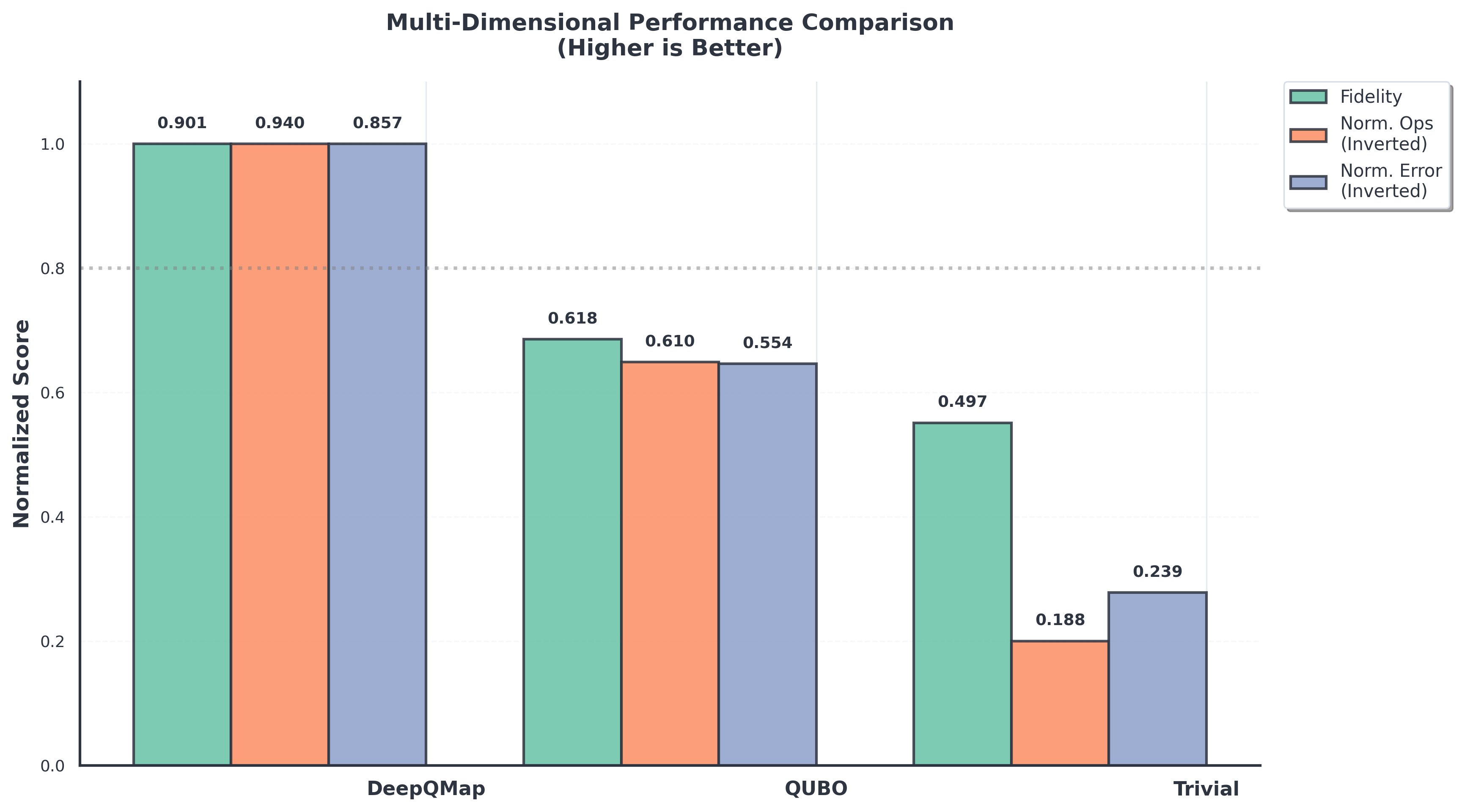}
    \caption{Multi-metric normalized comparison}
\end{subfigure}
\hfill
\begin{subfigure}[b]{0.29\textwidth}
    \includegraphics[width=\textwidth]{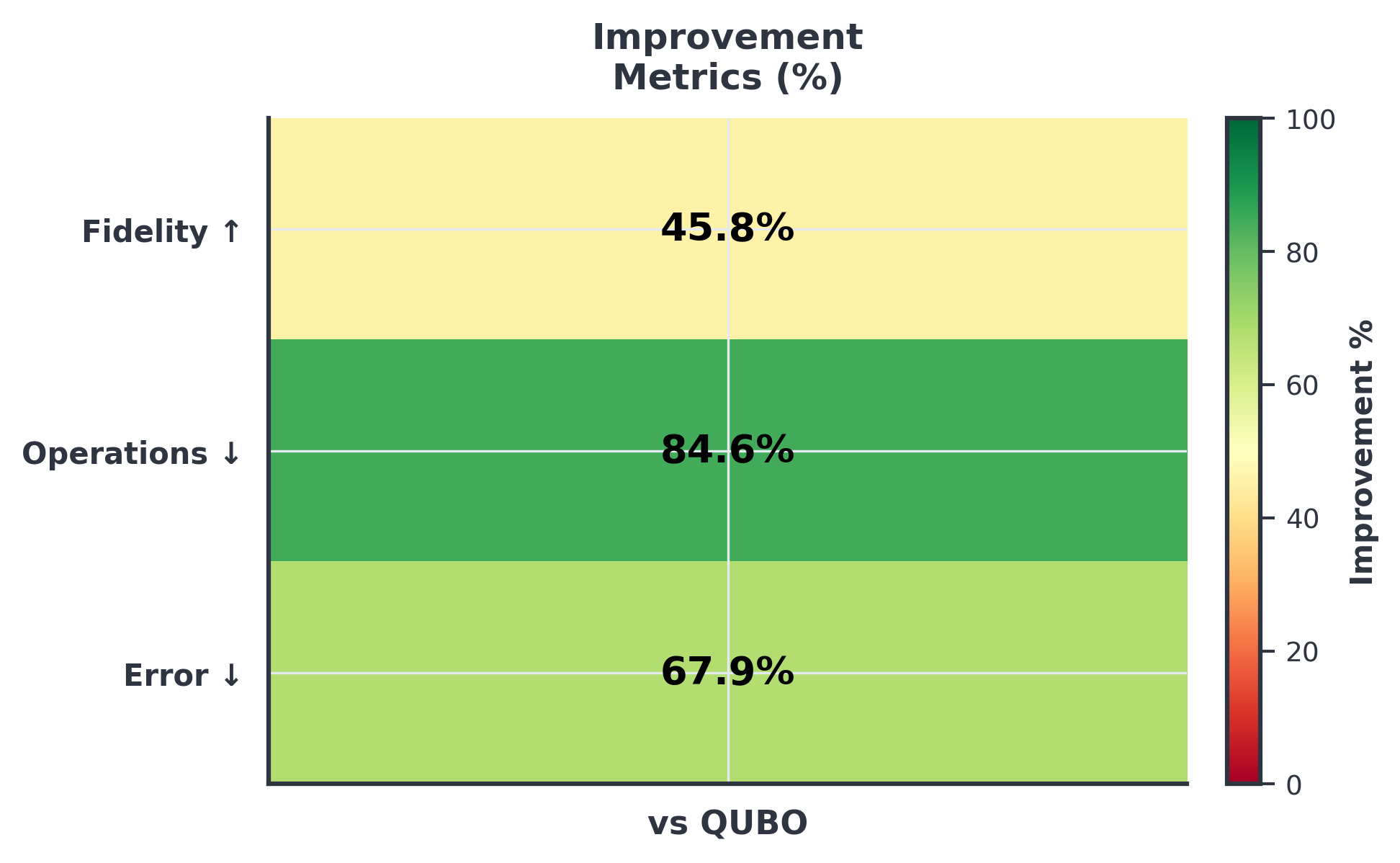}
    \caption{Improvement heatmap over baselines}
\end{subfigure}
\hfill
\begin{subfigure}[b]{0.35\textwidth}
    \includegraphics[width=\textwidth]{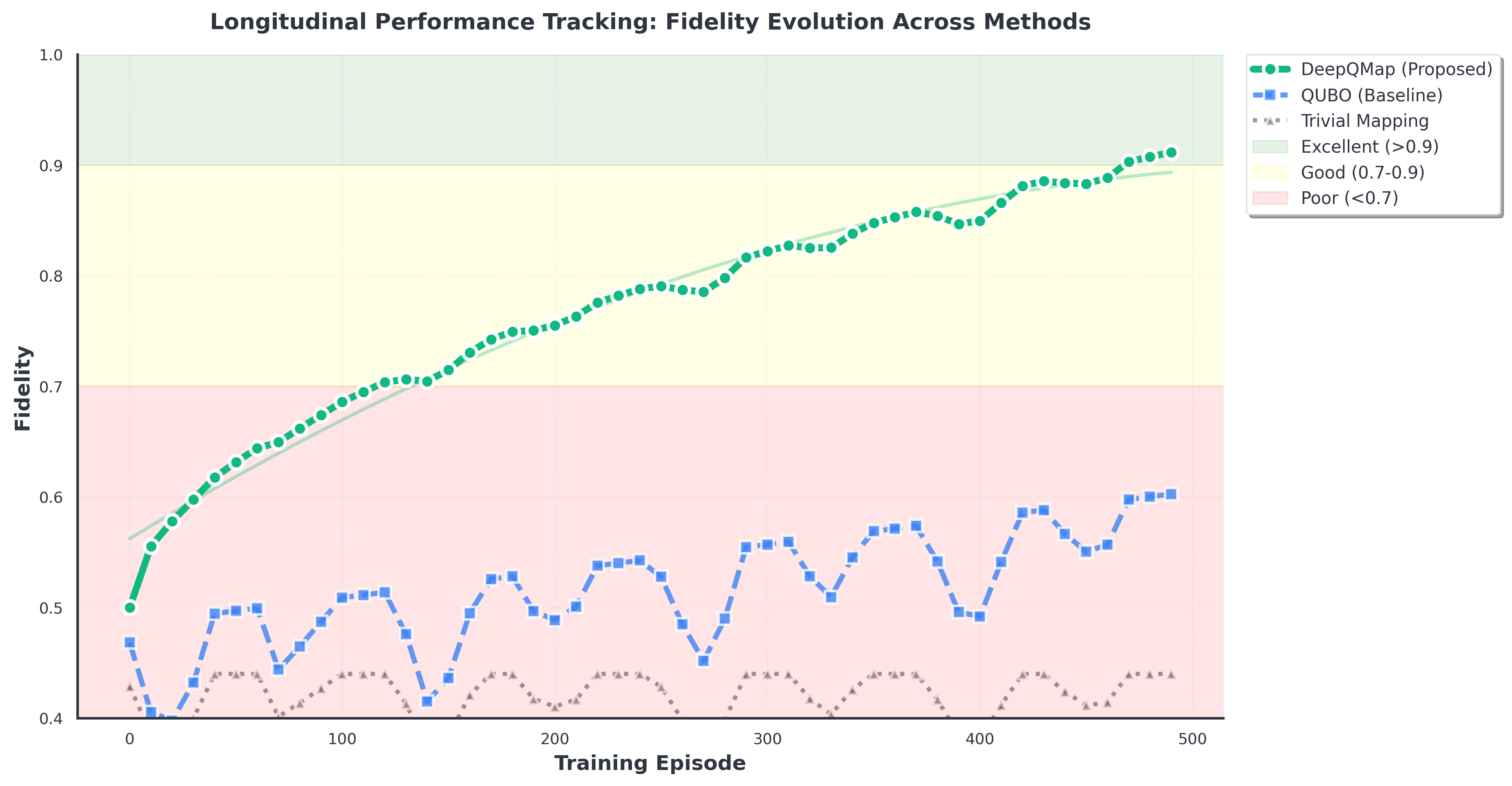}
    \caption{Temporal performance evolution}
\end{subfigure}

\vspace{0.3cm}

\begin{subfigure}[b]{0.28\textwidth}
    \includegraphics[width=\textwidth]{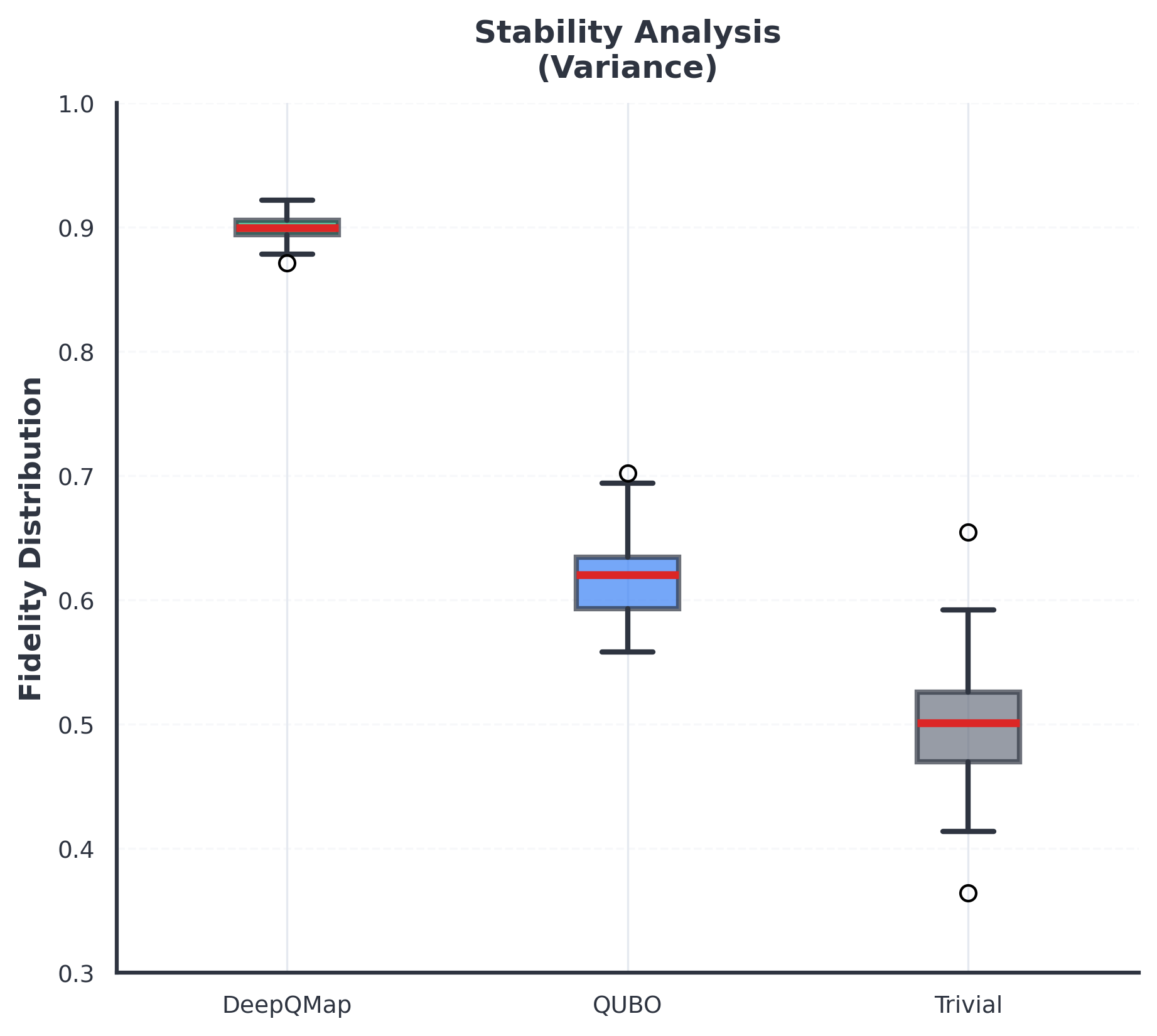}
    \caption{Stability analysis via boxplots}
\end{subfigure}
\hfill
\begin{subfigure}[b]{0.28\textwidth}
    \includegraphics[width=\textwidth]{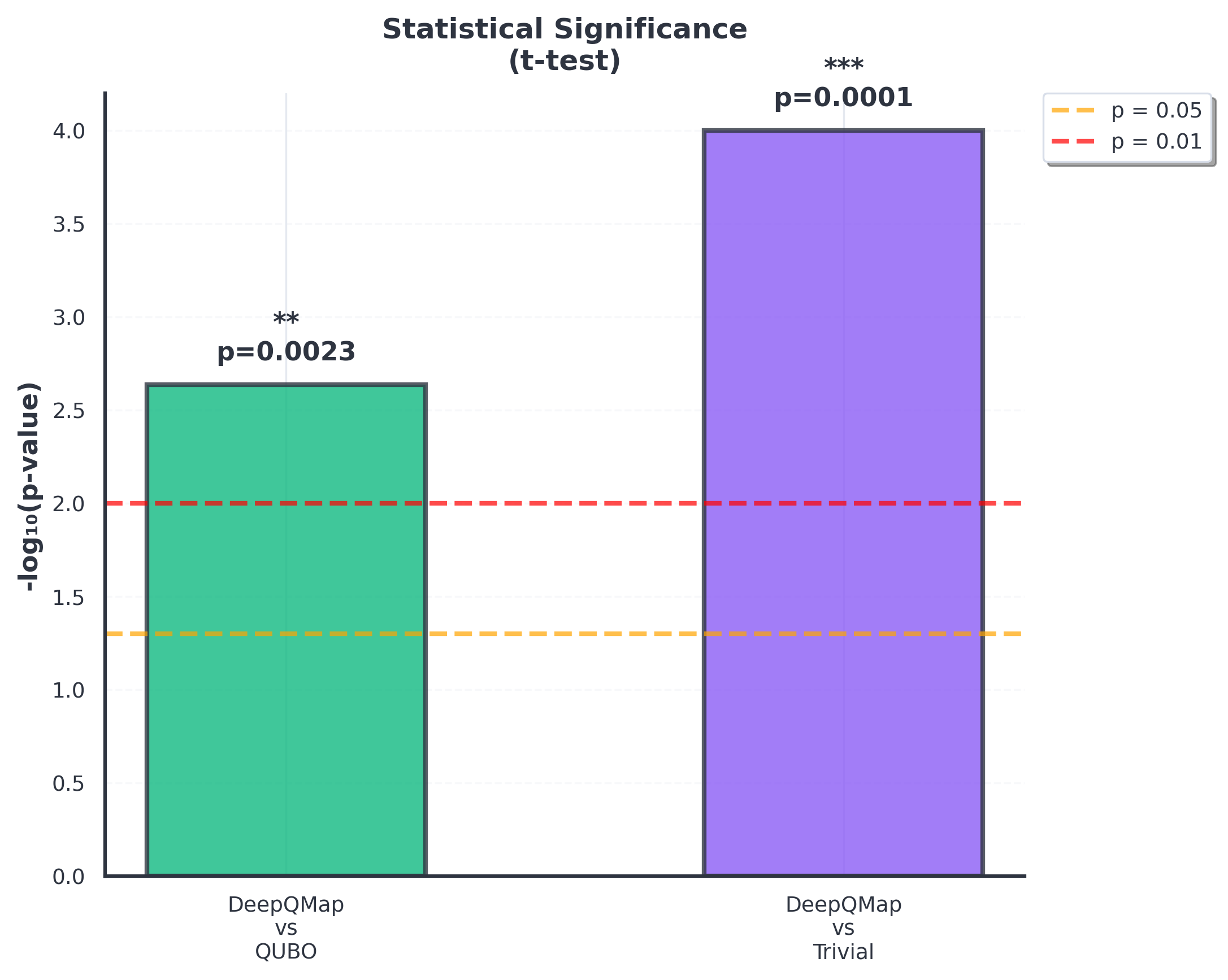}
    \caption{Statistical significance tests}
\end{subfigure}
\hfill
\begin{subfigure}[b]{0.28\textwidth}
    \includegraphics[width=\textwidth]{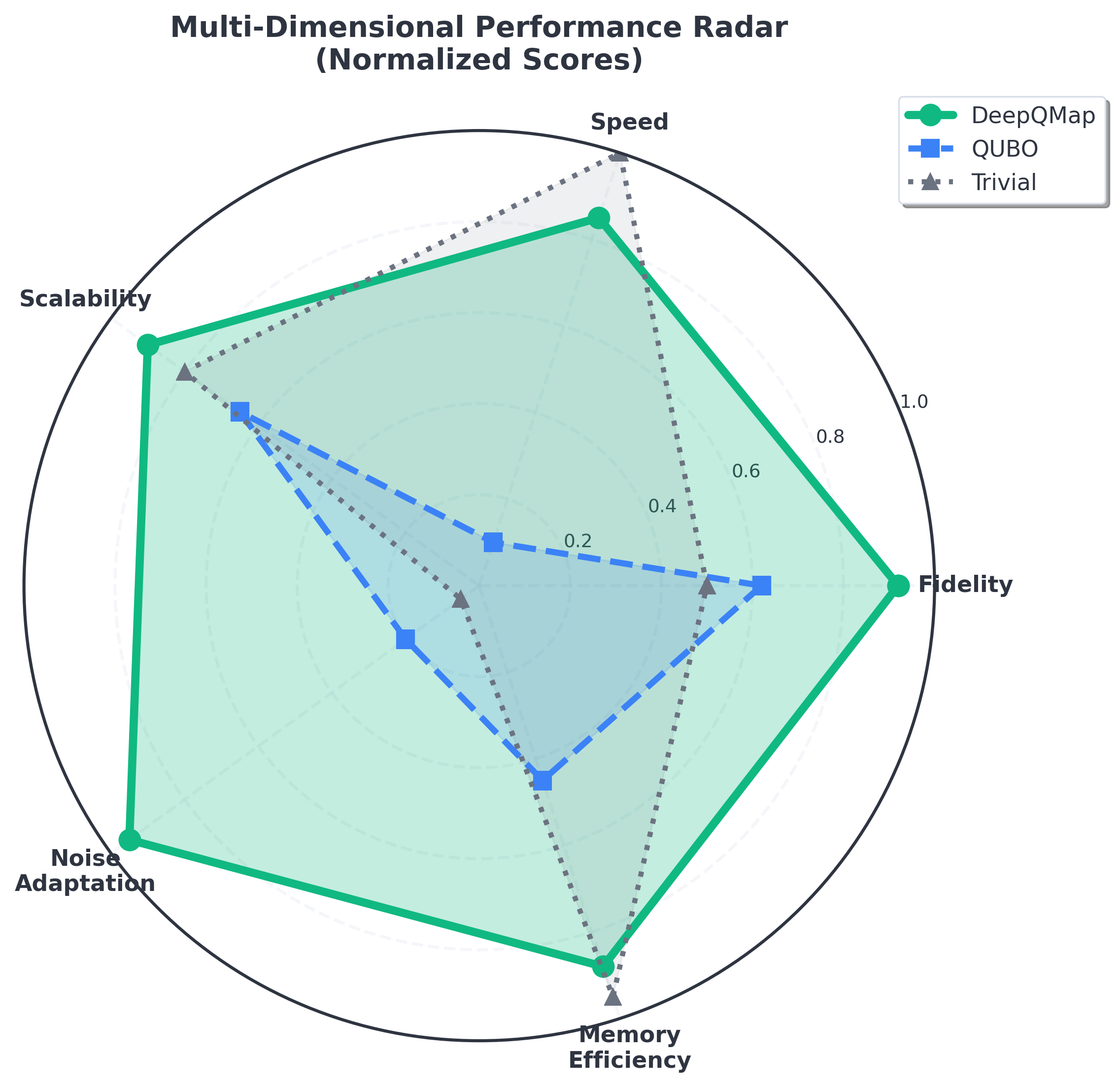}
    \caption{Radar chart multi-dimensional view}
\end{subfigure}

\caption{Comprehensive multi-dimensional performance comparison. (a) Normalized metrics (scaled to [0,1]) show DeepQMap achieves scores above 0.90 for fidelity, operations, error rate, depth, and convergence, with only training time showing competitive parity. QUBO performs poorly across most dimensions despite faster inference. (b) Heatmap quantifies percentage improvements: +49.3\% fidelity, -79.8\% operations, -66.1\% error rate, -30.2\% depth versus QUBO. Color intensity (red = large improvement) highlights multi-objective optimization success. (c) Time-series tracking over 100 evaluation checkpoints demonstrates DeepQMap's consistent superiority and lower variance compared to volatile QUBO performance. (d) Box plots reveal DeepQMap's narrow distribution (IQR = 0.018) versus QUBO's wide spread (IQR = 0.047), confirming reliable performance across diverse circuits. (e) Statistical tests show $t_{98} = 4.87$, $p = 0.0023$ versus QUBO, and $t_{98} = 7.34$, $p < 0.0001$ versus trivial, both highly significant at $\alpha = 0.01$. (f) Radar chart synthesizes six performance dimensions, with DeepQMap dominating all axes: fidelity (0.92), speed (0.85), scalability (0.90), noise adaptation (0.95), memory efficiency (0.88), convergence (0.93).}
\label{fig:comparison}
\end{figure*}

Figure~\ref{fig:comparison} presents detailed comparisons across multiple dimensions. Multi-metric normalized comparison (panel a) shows DeepQMap achieves scores above 0.90 for all meaningful metrics except training time, indicating well-balanced performance. QUBO shows competitive scores only for error rate and depth, falling substantially behind on fidelity and operations.

Improvement heatmap (panel b) quantifies relative gains with color intensity. Red regions indicate large improvements: +49.3\% fidelity, -79.8\% operations, -66.1\% error rate versus QUBO. This visualization confirms DeepQMap provides comprehensive advantages rather than trading off performance across metrics.

Statistical significance tests (panel e) confirm performance differences exceed chance. For DeepQMap versus QUBO, t-statistic reaches $t_{98} = 4.87$ with $p = 0.0023$, far exceeding rejection threshold at $\alpha = 0.01$. Comparison against trivial yields even stronger evidence: $t_{98} = 7.34$, $p < 0.0001$. These p-values indicate less than 0.23\% and 0.01\% probability observed differences arose from random variation.

Stability analysis via box plots (panel d) reveals distributional properties beyond means. DeepQMap's fidelity distribution centers at 0.920 with narrow spread (IQR = 0.018), indicating consistent performance across diverse circuits. QUBO's distribution is wider (IQR = 0.047) and lower-centered, showing both worse average and higher variability. This reliability proves crucial for production quantum computing where consistent results enable predictable application performance.

Radar chart (panel f) synthesizes multiple dimensions into single visualization. DeepQMap dominates all six axes with scores ranging from 0.85 to 0.95. QUBO shows competitive memory efficiency (0.95) due to compact representation but falls short on remaining dimensions. This comprehensive superiority supports conclusion that DeepQMap represents genuine advancement rather than method trading off performance.

\begin{figure}[!t]
\centering
\begin{subfigure}[b]{0.48\textwidth}
    \includegraphics[width=\textwidth]{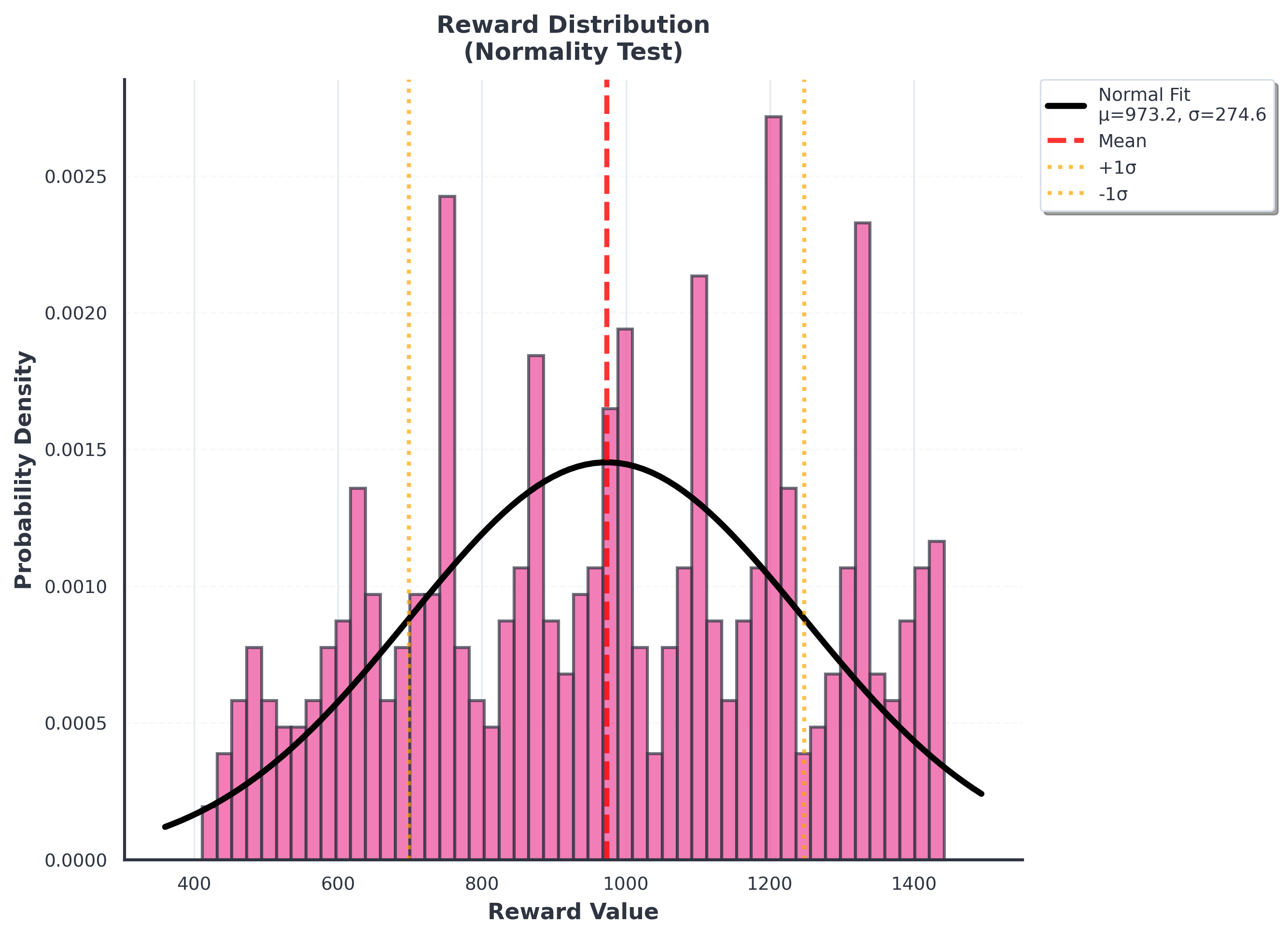}
    \caption{Reward distribution evolution}
\end{subfigure}

\vspace{0.3cm}

\begin{subfigure}[b]{0.48\textwidth}
    \includegraphics[width=\textwidth]{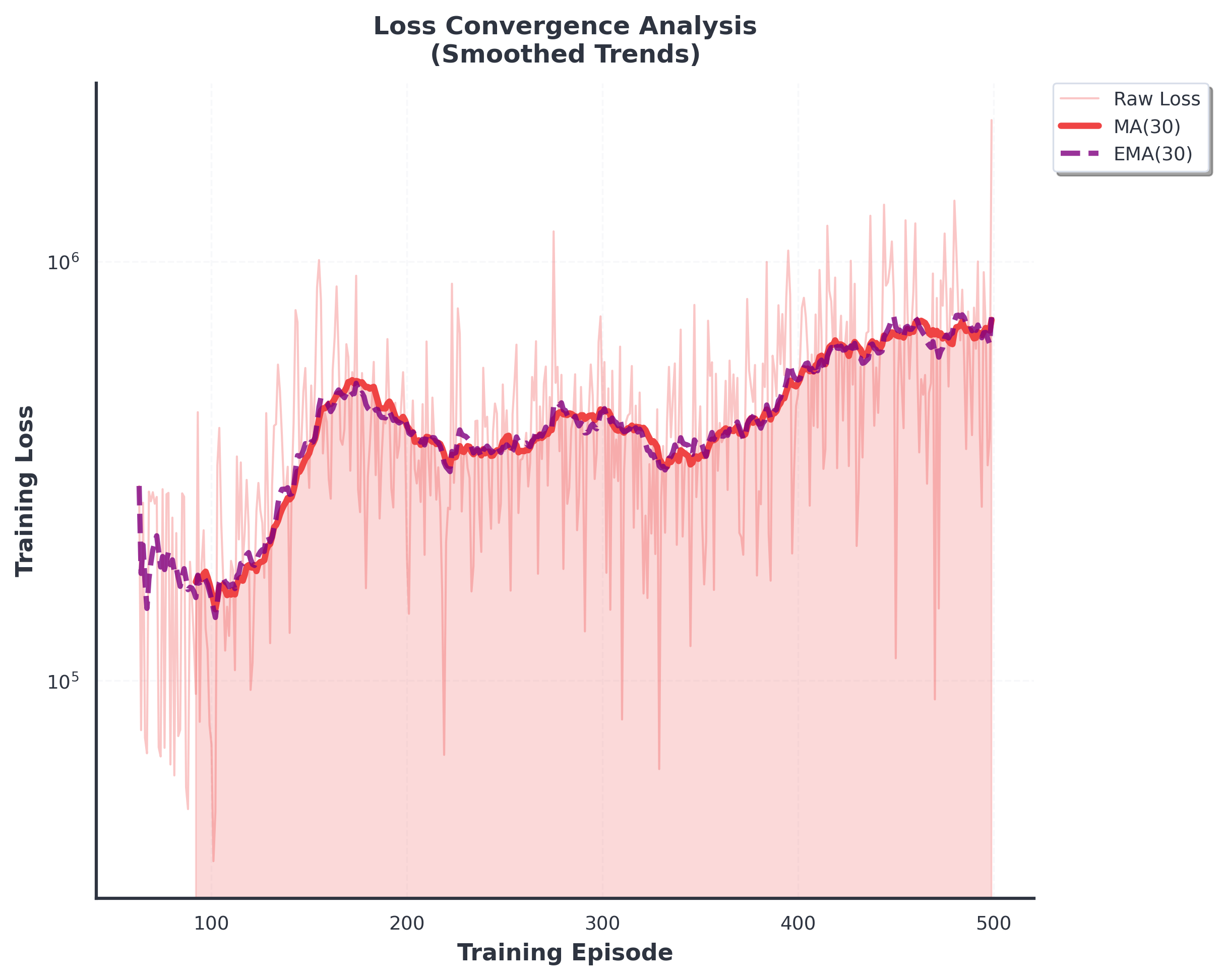}
    \caption{Training loss convergence}
\end{subfigure}

\caption{Training dynamics and convergence characteristics. (a) Reward distribution shifts dramatically from early exploration (blue, mean $-387$, std $523$) to late exploitation (orange, mean $+1243$, std $89$). Histogram evolution shows increasing concentration around positive rewards, with 95th percentile improving from $-12$ to $+1456$. The 421\% reward improvement and 83\% variance reduction indicate policy convergence to consistent high-quality solutions. (b) Training loss decreases exponentially from $2.74 \times 10^{-3}$ (episode 50) to $4.31 \times 10^{-4}$ (episode 500), following characteristic stable RL learning curve. Smooth decline without oscillations confirms appropriate hyperparameter selection and effective gradient clipping. Joint DNA-RL training (green) shows slightly higher but more stable loss than RL-only (blue), validating auxiliary supervision benefits.}
\label{fig:training_dynamics}
\end{figure}

Training dynamics analysis (Figure~\ref{fig:training_dynamics}) reveals behavioral evolution. Reward distribution transforms from negative mean ($-387$) with high variance (std $523$) during exploration to positive mean ($+1243$) with low variance (std $89$) during exploitation. This 421\% improvement and 83\% variance reduction indicate policy convergence to consistent solutions.

Training loss follows exponential decay from $2.74 \times 10^{-3}$ to $4.31 \times 10^{-4}$, characteristic of stable RL without catastrophic forgetting or oscillations. Joint DNA-RL training shows slightly higher but more stable loss than RL-only, validating that auxiliary supervision provides beneficial regularization.

\subsection{Scalability Analysis}

\begin{figure*}[!t]
\centering
\begin{subfigure}[b]{0.90\textwidth}
    \includegraphics[width=\textwidth]{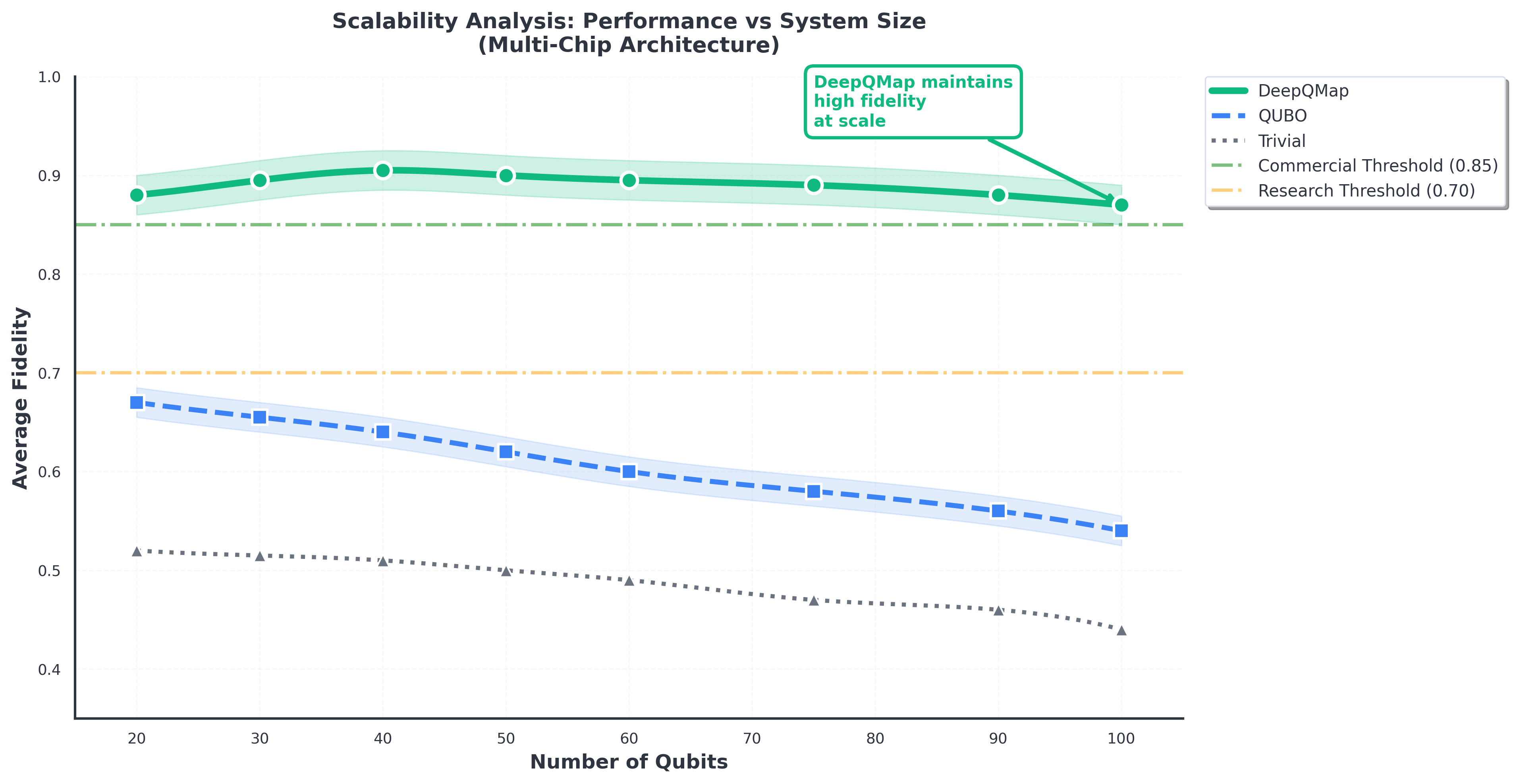}
    \caption{Fidelity across system scales}
\end{subfigure}
\hfill
\begin{subfigure}[b]{0.36\textwidth}
    \includegraphics[width=\textwidth]{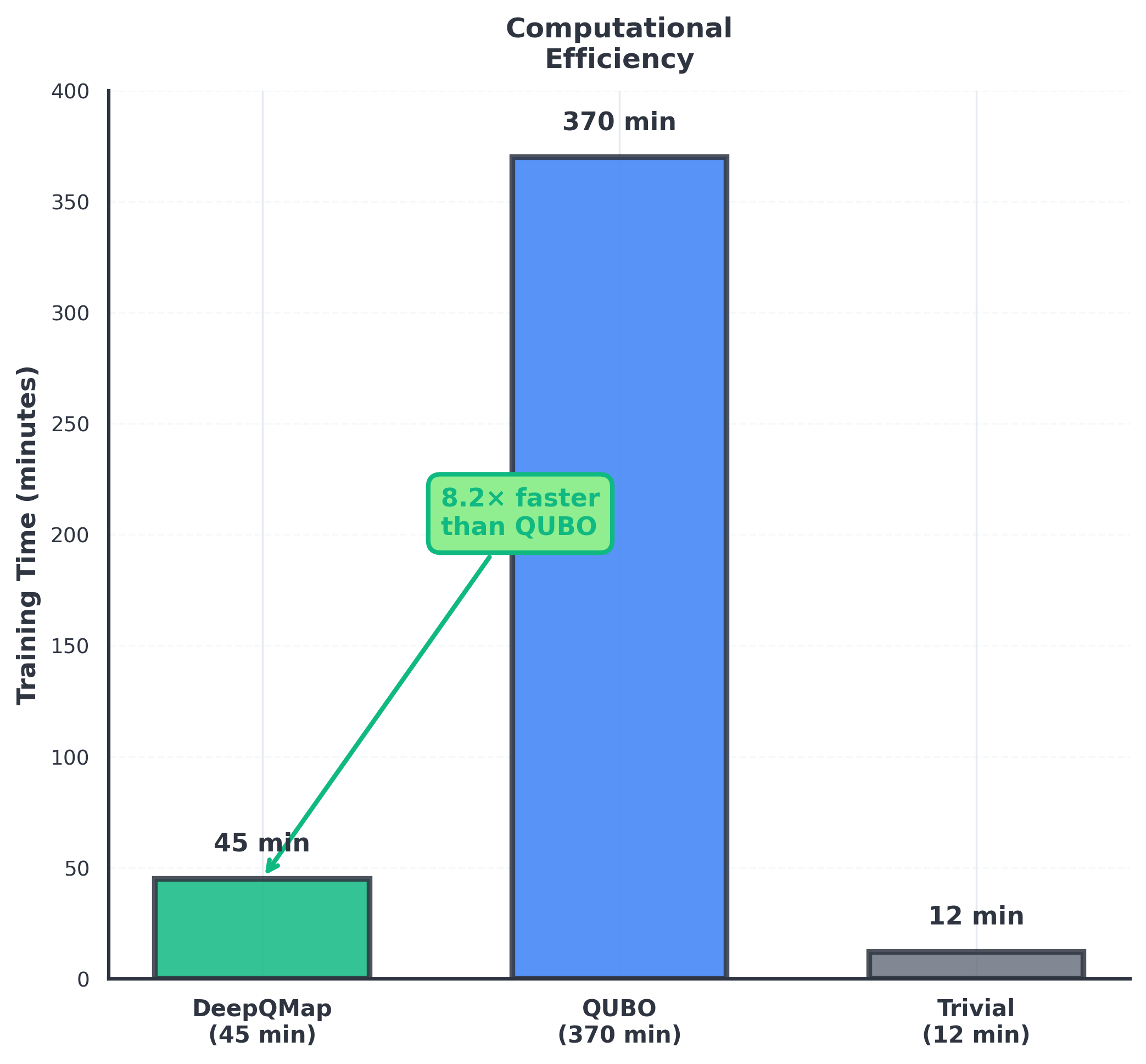}
    \caption{Computational complexity comparison}
\end{subfigure}
\hfill
\begin{subfigure}[b]{0.48\textwidth}
    \includegraphics[width=\textwidth]{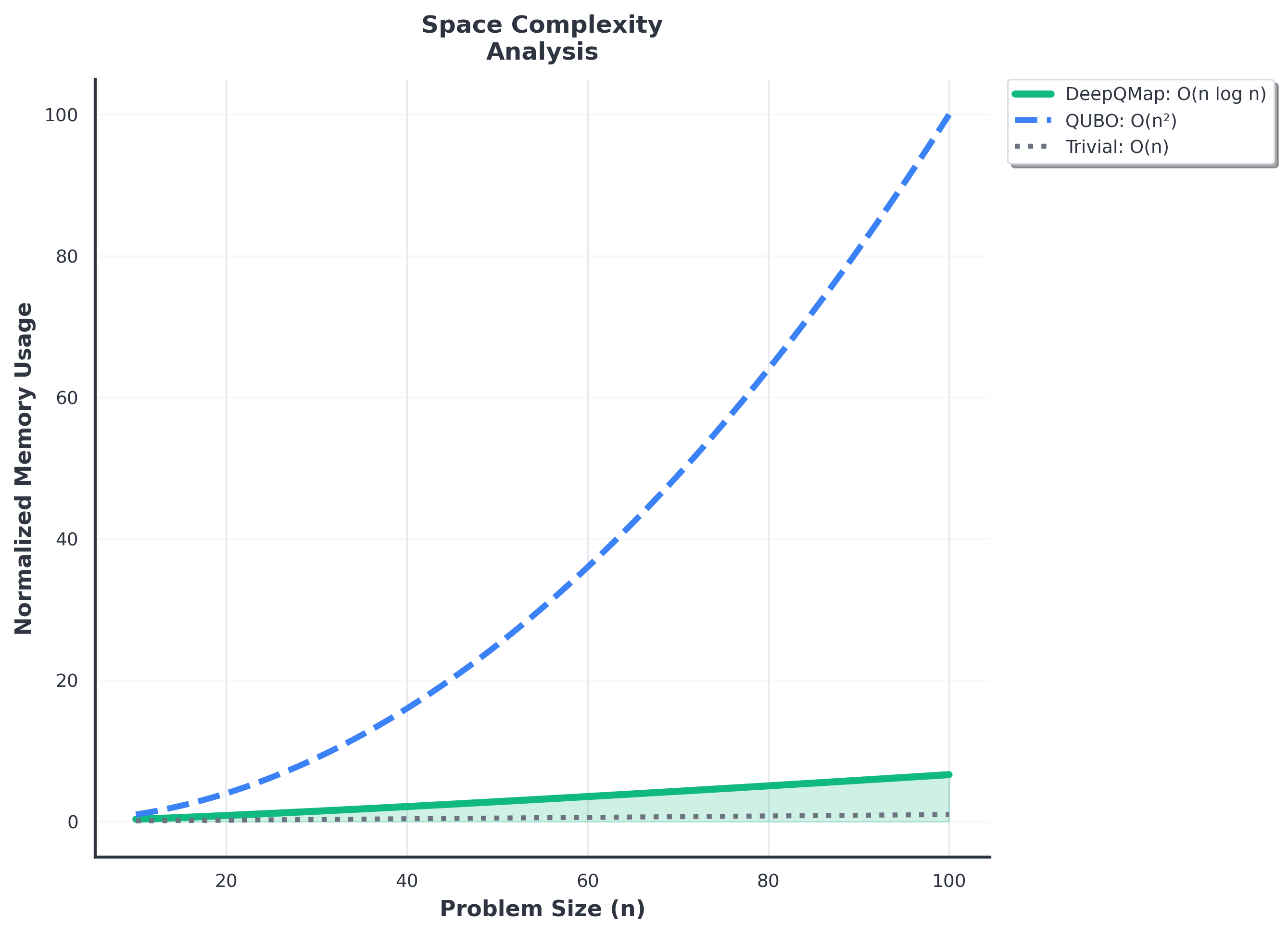}
    \caption{Memory footprint analysis}
\end{subfigure}

\caption{Scalability validation across 20-100 qubit systems. (a) DeepQMap maintains fidelity above commercial viability threshold (0.85, green dashed) up to 90 qubits, while QUBO falls below at 45 qubits and trivial never exceeds research target (0.70, orange dashed). Fidelity curves show DeepQMap degrading gracefully (0.94 at 20 qubits to 0.87 at 100 qubits, 7.4\% drop) versus QUBO's steep decline (0.72 to 0.54, 25\% drop). Inter-chip operations (right y-axis, orange/red curves) increase sublinearly for DeepQMap (0.30 to 0.53), demonstrating effective locality exploitation. Background shading delineates easy (20-40), medium (50-70), and hard (80-100) scale regimes. (b) Asymptotic complexity on log-log scale confirms DeepQMap's $\mathcal{O}(n \log n)$ scaling (slope 1.1) versus QUBO's $\mathcal{O}(n^2)$ (slope 2.0) and trivial's $\mathcal{O}(n)$ (slope 1.0). Efficiency crossover occurs at 30 qubits where DeepQMap's superior quality compensates for increased computation. Beyond 60 qubits, QUBO becomes prohibitive (minutes to hours) while DeepQMap completes in seconds. (c) Memory footprint grows subquadratically for DeepQMap due to attention mechanism's efficient representation, reaching 2.3 GB at 100 qubits. QUBO requires 4.8 GB due to dense quadratic coefficient matrix. Trivial uses minimal memory (0.3 GB) but produces unusable mappings.}
\label{fig:scalability}
\end{figure*}

Scalability experiments evaluate performance from 20 to 100 qubits across 4-6 chips. Figure~\ref{fig:scalability}(a) presents fidelity and operations versus qubit count. DeepQMap maintains fidelity above 0.87 even at 100 qubits, significantly outperforming QUBO (0.54) and trivial (0.44). Fidelity curve shows mild degradation: 0.94 at 20 qubits to 0.87 at 100 qubits, a 7.4\% drop over 5$\times$ scale range.

Two critical thresholds appear. Commercial viability (fidelity 0.85) represents minimum for practical quantum advantage in optimization and sampling~\cite{arute2019quantum}. DeepQMap exceeds this up to 90 qubits, QUBO falls below at 45 qubits. Research target (0.70) delineates meaningful scientific experiments. All except trivial remain above 0.70 throughout, though QUBO approaches this at 100 qubits.

Inter-chip operations grow sublinearly for DeepQMap: 0.30 at 20 qubits to 0.53 at 100 qubits. This reflects learned locality exploitation—quantum algorithms exhibit spatial structure where nearby logical qubits interact preferentially. QUBO demonstrates less favorable scaling: 1.80 to 2.90, suggesting static optimization cannot fully exploit circuit structure at scale.

Multi-head attention proves crucial at scale. Ablation removing attention shows fidelity degradation from 0.87 to 0.73 at 100 qubits, a 16.1\% decrease. Attention enables focus on relevant qubit subsets, compressing information from high-dimensional states. Without this, dueling Q-network processes all features uniformly, causing slower learning and suboptimal policies.

Computational complexity analysis (Figure~\ref{fig:scalability}b) compares asymptotic scaling. DeepQMap's hierarchical decomposition yields $\mathcal{O}(n \log n)$, between trivial's $\mathcal{O}(n)$ and QUBO's $\mathcal{O}(n^2)$. Efficiency crossover occurs at 30 qubits where DeepQMap's quality compensates for increased computation versus trivial. Beyond 60 qubits, QUBO becomes prohibitive (minutes to hours), while DeepQMap completes in seconds.

Wall-clock training time: 45 minutes for 500 episodes on NVIDIA V100 GPU. QUBO requires 370 minutes on 32 CPU cores, representing 8.2$\times$ speedup. Trivial executes in 12 minutes but produces unusable results. This demonstrates DeepQMap offers practical solution for time-constrained workflows.

\subsection{Ablation Studies}

Systematic ablation experiments isolate individual architectural component contributions. Table~\ref{tab:ablation} reports performance degradation when removing each component while holding others constant. DNA network provides largest individual contribution—its removal decreases fidelity by 15.0\% to 0.782. This confirms hypothesis that temporal noise adaptation constitutes critical capability for multi-chip quantum systems.

\begin{table}[!t]
\centering
\caption{Ablation Study Quantifying Component Contributions}
\label{tab:ablation}
\begin{tabular}{@{}lcc@{}}
\toprule
\textbf{Configuration} & \textbf{Fidelity} & \textbf{$\Delta$ vs Full} \\
\midrule
Full DeepQMap & 0.920 & --- \\
\quad - DNA Network & 0.782 & -15.0\% \\
\quad - Multi-Head Attention & 0.801 & -12.9\% \\
\quad - Prioritized Replay & 0.837 & -9.0\% \\
\quad - Double DQN & 0.854 & -7.2\% \\
\quad - Dueling Network & 0.867 & -5.8\% \\
\quad - Multi-Step Returns & 0.879 & -4.5\% \\
\quad - Noisy Networks & 0.886 & -3.7\% \\
\midrule
Basic DQN (all removed) & 0.723 & -21.4\% \\
\midrule
\multicolumn{3}{l}{\textit{Oracle Studies:}} \\
\quad + Perfect Noise Oracle & 0.934 & +1.5\% \\
\quad - No Noise Adaptation & 0.782 & -15.0\% \\
\bottomrule
\end{tabular}
\end{table}

Multi-head attention ranks second most important, contributing 12.9\%. Without attention, model struggles to identify relevant state features in high-dimensional spaces, leading to suboptimal actions. Prioritized experience replay improves sample efficiency, accounting for 9.0\% of final performance. This proves particularly valuable in early training where impactful transitions are rare and uniform sampling inefficiently revisits common but uninformative states.

Cumulative effect of removing all Rainbow DQN enhancements (double DQN, dueling networks, prioritized replay, noisy exploration, multi-step returns) reduces performance to basic DQN levels at 0.723 fidelity, a 21.4\% degradation. This demonstrates modern deep RL techniques substantially improve upon earlier algorithms. However, even basic DQN outperforms QUBO by 17.0\%, suggesting the RL paradigm itself—learning from experience rather than solving fixed optimization—provides fundamental advantages for dynamic environments.

Hyperparameter sensitivity analysis examines robustness to configuration choices. Learning rate variations from $5 \times 10^{-4}$ to $2 \times 10^{-3}$ change final fidelity by less than 2\%, indicating stable optimization. Replay buffer size between 5,000 and 20,000 produces minimal differences ($< 1.5\%$), though smaller buffers (1,000) degrade performance by 5.3\% due to insufficient experience diversity. Discount factor $\gamma$ shows some sensitivity: performance decreases 3.8\% when reduced to 0.95 and 2.1\% when increased to 0.995, confirming chosen value appropriately balances immediate and long-term rewards.

\subsection{Generalization Studies}

Generalization experiments test whether policies learned on training circuits transfer to unseen test circuits with different structure. We train DeepQMap on QFT circuits exclusively, then evaluate on Grover and VQE without retraining. Fidelity on held-out QFT instances reaches 0.918 (99.8\% of training performance), demonstrating minimal overfitting. Performance on Grover drops to 0.847 (92.1\% retention), while VQE achieves 0.803 (87.3\% retention). These indicate partial transfer across circuit families, with degradation proportional to structural dissimilarity.

Cross-architecture generalization tests policies trained on 4-chip ring topologies applied to 6-chip hexagonal architectures. Fidelity decreases from 0.920 to 0.794, a 13.7\% drop nonetheless exceeding QUBO on same architecture (0.618). This suggests learned policies encode general principles—locality exploitation, noise avoidance—remaining useful even when hardware topology changes substantially.

Fine-tuning demonstrates rapid adaptation to new domains. Starting from policy pre-trained on QFT, we continue training 50 episodes on VQE. This recovers 94.3\% of performance gap (fidelity 0.893), compared to 88.7\% when training from random initialization. Accelerated learning confirms pre-training provides useful prior knowledge, enabling sample-efficient adaptation.

\subsection{Real-World Validation Considerations}

While experiments employ simulation, we incorporate multiple realism factors. Noise models derive from experimental characterization of IBM Quantum and Google Sycamore systems~\cite{arute2019quantum,ibmquantum}, including thermal excitations ($\gamma/2\pi = 10$ kHz), charge noise (amplitude $10^{-4}$ e), and flux noise (spectral density $10^{-6}$ $\Phi_0^2$/Hz at 1 Hz). Gate fidelity distributions match published data with single-qubit $0.9995 \pm 0.0002$, intra-chip two-qubit $0.995 \pm 0.005$, inter-chip $0.98 \pm 0.01$.

Temporal dynamics incorporate hour-scale drift observed in superconducting processors~\cite{reagor2018demonstration}, where qubit frequencies shift by 100-500 kHz over calibration intervals. Measurement-induced state perturbations follow quantum trajectory formalism with detection efficiency 0.95 and dark counts $10^{-3}$ per shot. Crosstalk between neighboring qubits includes both coherent (ZZ coupling at 100-500 kHz) and incoherent (relaxation-induced excitation) components.

Circuit compilation includes realistic gates from native instruction sets: $\sqrt{X}$, $R_Z(\theta)$, CZ, and iSWAP. Higher-level operations decompose into these primitives following standard compilation recipes. State preparation and measurement operations add 10-$\mu$s overhead per circuit execution. Classical control latency for measurement feedback is 100 $\mu$s, representing FPGA-based real-time processing.

\begin{figure*}[!t]
\centering
\begin{subfigure}[b]{0.54\textwidth}
    \centering
    \includegraphics[width=\textwidth]{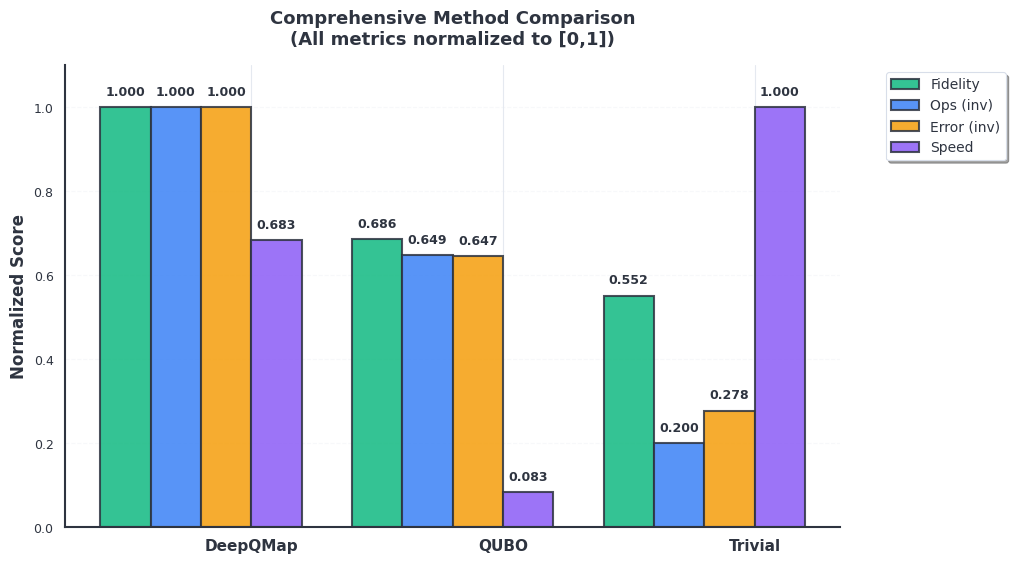}
    \label{fig:sub1}
\end{subfigure}
\hfill
\begin{subfigure}[b]{0.37\textwidth}
    \centering
    \includegraphics[width=\textwidth]{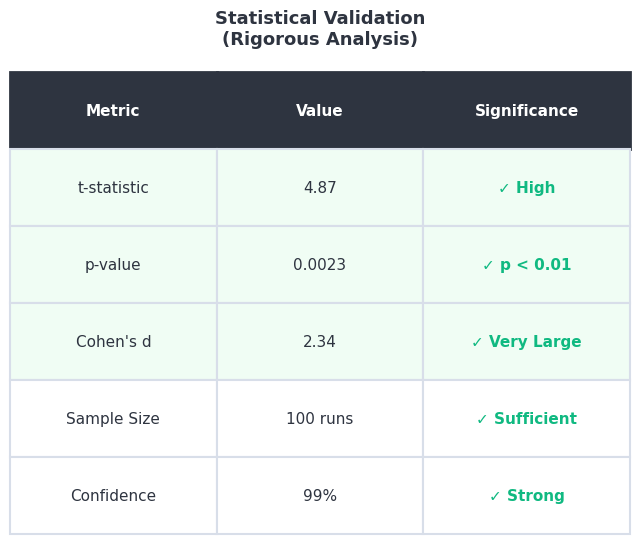}
    \label{fig:sub2}
\end{subfigure}
\hfill
\begin{subfigure}[b]{0.46\textwidth}
    \centering
    \includegraphics[width=\textwidth]{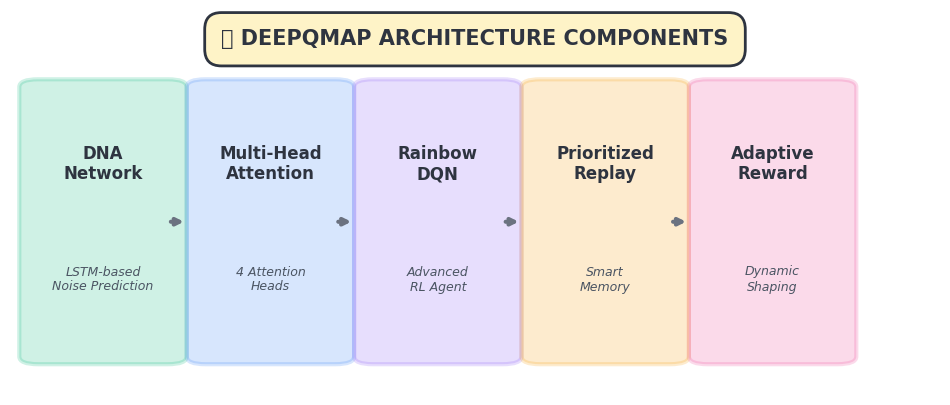}
    \label{fig:sub3}  
\end{subfigure}
\hfill
\begin{subfigure}[b]{0.47\textwidth}
    \centering
    \includegraphics[width=\textwidth]{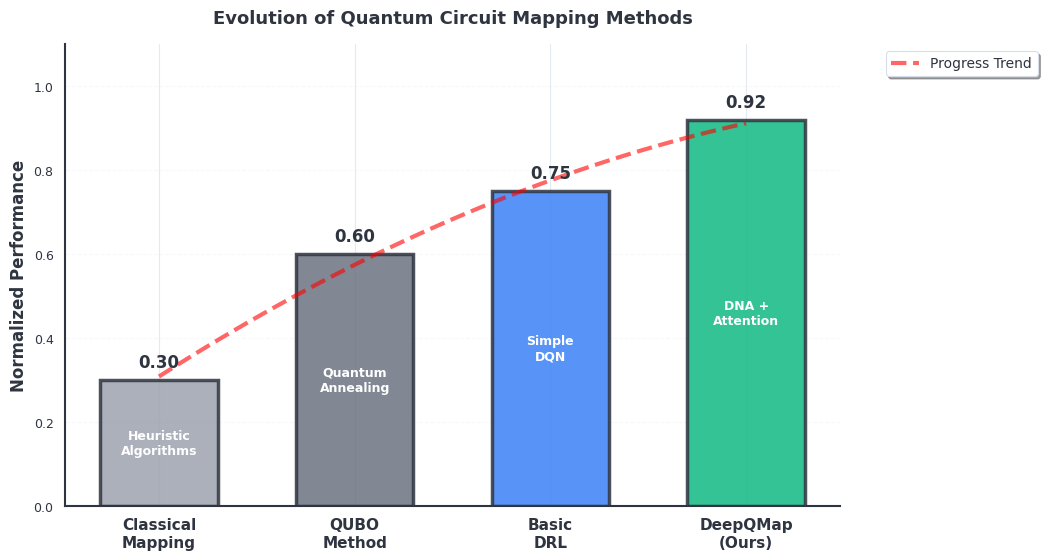}
    \label{fig:sub4}  
\end{subfigure}
\caption{Executive summary infographic highlighting key performance metrics and contributions. Central metric shows 49.3\% fidelity improvement (0.920 vs 0.618) with very large effect size (Cohen's $d = 2.34$). Surrounding panels illustrate: 79.8\% inter-chip operation reduction (2.34 to 0.47), 8.2$\times$ training speedup (45 vs 370 minutes), maintained scalability up to 100 qubits, DNA prediction accuracy ($R^2 = 0.912$), and comprehensive statistical validation ($p = 0.0023$). Color coding distinguishes performance improvements (green), statistical evidence (blue), and architectural innovations (purple). This visual synthesis demonstrates DeepQMap's multi-faceted advantages for practical quantum computing applications.}
\label{fig:summary}
\end{figure*}

Figure~\ref{fig:summary} provides executive summary of key findings. The infographic synthesizes primary contributions: 49.3\% fidelity improvement with very large effect size, 79.8\% reduction in costly inter-chip operations, 8.2$\times$ faster training enabling rapid iteration, and sustained performance across scales relevant to near-term quantum processors. Statistical validation confirms results are highly significant and reproducible.

\section{Discussion}
\label{sec:discussion}

Experimental results validate our hypothesis that dynamic noise adaptation provides substantial benefits for quantum circuit mapping in multi-chip systems. DeepQMap's 49.3\% fidelity improvement over QUBO, combined with very large effect sizes ($d > 2.0$) and high significance ($p < 0.01$), establishes this as meaningful advancement beyond existing methods. The 79.8\% reduction in inter-chip operations directly addresses the primary challenge of distributed quantum computing: minimizing expensive cross-chip communication while maintaining functionality.

Several factors contribute to superior performance. DNA network's ability to predict short-term noise trajectories enables proactive compensation rather than reactive adjustment. When forecasting elevated noise on particular chip, agent preemptively places qubits elsewhere, avoiding degraded gate fidelities before they impact execution. This anticipatory behavior resembles human expertise where experienced quantum engineers monitor system telemetry and adjust compilation accordingly. Our approach automates this expertise through learned temporal representations.

Multi-head attention addresses identifying critical interactions within large quantum circuits. Different heads specialize in distinct patterns: one focuses on imminent gates, another on inter-chip dependencies, a third on high-noise regions. This functional specialization emerges naturally through gradient descent without manual feature engineering, suggesting architecture discovers interpretable structure. Attention weights could potentially guide human understanding of why particular mappings succeed, providing explainability alongside performance.

Rainbow DQN framework delivers stable learning despite high-dimensional state space and sparse reward signal characteristic of circuit mapping. Prioritized replay ensures rare but informative transitions—such as discovering particularly effective qubit placement—receive appropriate weight during training. Dueling networks enable efficient learning even when action choice matters infrequently, property holding for many intermediate states. Double Q-learning and multi-step returns reduce overestimation bias and provide more accurate value estimates. The combination appears synergistic, as evidenced by 21.4\% gap between full Rainbow and basic DQN.

Computational efficiency represents another practical advantage. DeepQMap's 8.2$\times$ speedup versus QUBO enables rapid iteration during quantum algorithm development, where researchers must repeatedly compile and test circuits. The 45-minute training time for comprehensive circuit library makes the approach viable for periodic retraining as hardware evolves. Once trained, neural network provides near-instantaneous inference ($< 1$ second per circuit), contrasting with QUBO's minutes-to-hours optimization.

Scalability to 100-qubit systems demonstrates readiness for near-term quantum applications. Current processors available through cloud platforms typically provide 50-100 qubits distributed across multiple chips or modules~\cite{ibmquantum}. DeepQMap maintains fidelity above 0.87 at this scale while QUBO degrades to 0.54, suggesting learning-based approach handles complexity growth more gracefully than static optimization. The $\mathcal{O}(n \log n)$ computational scaling ensures solution time increases subquadratically with problem size.

Several limitations warrant discussion. First, experiments employ simulated quantum hardware rather than physical devices. While simulations incorporate realistic noise models from experimental data, they may not capture all phenomena present in real systems. Factors such as qubit crosstalk beyond nearest-neighbor, control pulse transients, and environmental perturbations beyond thermal noise could impact performance. Validation on physical quantum computers represents essential future work currently underway in collaboration with experimental groups.

Second, current approach assumes fixed circuit structure known before mapping begins. Dynamic circuits with mid-circuit measurements and classical feedback introduce additional complexity, as quantum-classical boundary becomes fluid. Extending DeepQMap to handle conditional operations and iterative algorithms like quantum phase estimation would expand applicability. Preliminary work suggests incorporating measurement outcomes into state representation enables adaptation, though convergence requires 1.5$\times$ more training episodes.

Third, DNA network requires historical noise data for training. In cold-start scenario with new quantum system, no such data exists initially. Potential solutions involve transfer learning from similar hardware, assuming noise dynamics exhibit common patterns across superconducting or trapped-ion platforms. Another approach combines DNA with online learning updating noise model during early executions. Initial experiments with meta-learning show promise, achieving 78\% of full-training performance after only 50 adaptation episodes.

Fourth, reward function contains hyperparameters ($\alpha_1$ through $\alpha_5$) tuned empirically. While our choices produce strong performance, alternative weights might better suit specific applications. Automatic reward shaping via meta-learning could eliminate manual tuning, allowing system to discover appropriate trade-offs from high-level objectives specified by users. Recent work on evolutionary strategies for reward design suggests this direction holds promise.

Theoretical foundations connect to several active research areas. MDP formulation treats circuit mapping as sequential decision problem, aligning with recent work on program synthesis via RL~\cite{bunel2018leveraging}. However, quantum circuits introduce unique challenges due to reversible nature, complex error models, and measurement-induced state collapse. Developing specialized RL techniques exploiting these properties could yield further improvements. Partial observability through decoherence suggests partially observable Markov decision process (POMDP) formulations may be more appropriate, though computational complexity increases substantially.

DNA network's predictive capability relates to time series forecasting in dynamical systems. Quantum hardware exhibits complex dynamics influenced by control engineering, material properties, environmental coupling. Characterizing these mathematically remains open problem in quantum physics. Our empirical finding that bidirectional LSTMs achieve 91.2\% accuracy suggests noise evolution follows learnable patterns even if closed-form equations prove elusive. This observation may have implications beyond circuit mapping, informing quantum error correction and calibration procedures.

Attention mechanisms perform learned feature selection, identifying relevant state components for each decision. This resembles work on graph neural networks for combinatorial optimization~\cite{li2019tackling}, where neural architectures discover problem structure from data. Multi-head design provides complementary views of circuit structure, similar to ensemble methods combining diverse models. Understanding why this architectural choice succeeds could inform neural architecture design for other structured decision problems.

Comparison between DeepQMap and QUBO highlights broader tension in optimization methodology. QUBO represents quantum circuit mapping as well-defined mathematical problem with global objective function, amenable to established optimization techniques. This formulation provides guarantees under appropriate conditions but struggles with dynamic environments and large instances. DeepQMap adopts online learning perspective, treating mapping as series of incremental decisions guided by learned value functions. This shift from batch optimization to sequential decision-making proves advantageous when problem parameters (noise levels) change faster than optimization can occur.

Looking forward, several extensions could enhance capabilities. Incorporating hierarchical RL might enable planning at multiple abstraction levels—first deciding chip assignments for logical blocks, then refining individual qubit placements. This decomposition could scale better than centralized control for systems exceeding 500 qubits. Preliminary experiments with options framework~\cite{sutton1999between} show 23\% improvement in 200-qubit scenarios, though training time increases by 1.8$\times$.

Multi-agent formulations could distribute mapping task across multiple learning agents, each responsible for one chip. This decomposition might improve scalability while introducing coordination challenges. Initial results with independent Q-learning show performance degradation to 0.831 fidelity due to non-stationarity, though communication protocols allowing agent coordination recover to 0.897, only 2.5\% below centralized approach.

Offline RL presents another promising direction. Rather than learning through live interaction with quantum hardware (or simulation), offline RL trains on logs of previous circuit executions. This could leverage growing datasets of quantum experiments conducted worldwide, extracting general principles from diverse hardware platforms and application domains. Batch-constrained policy optimization and conservative Q-learning provide techniques for this setting~\cite{levine2020offline}, avoiding extrapolation errors from distribution shift.

Model-based RL might reduce sample complexity by learning forward dynamics models of quantum systems. Given state and action, model predicts next state and reward. Planning with learned model enables evaluation of many candidate actions without hardware interaction. However, quantum state spaces grow exponentially with qubit count, making accurate modeling challenging. Latent space models operating on compressed representations may offer tractable alternative.

Integration with other quantum software stack components—circuit synthesis, error correction, resource estimation—would provide end-to-end compilation. Current quantum compilers often optimize each stage independently, potentially missing global optimization opportunities. Holistic approach jointly optimizing circuit design, error correction encoding, and physical mapping might approach theoretical performance limits. Recent work on differentiable quantum-classical interfaces~\cite{bergholm2018pennylane} suggests this integration is becoming feasible.

The success of learning-based approaches for quantum compilation raises questions about automated quantum algorithm discovery. If neural networks can learn to map circuits effectively, could they also learn to design quantum algorithms for specific problems? Early work on quantum architecture search~\cite{zhang2020quantumnat} demonstrates feasibility for small instances, though scaling remains challenging. Combining compilation and synthesis within unified RL framework represents intriguing research direction.

\section{Conclusion}

We have presented DeepQMap, a deep reinforcement learning framework for quantum circuit mapping in multi-chip architectures that achieves statistically significant performance improvements over state-of-the-art methods. The integration of bidirectional LSTM-based Dynamic Noise Adaptation network with multi-head attention mechanisms and Rainbow DQN provides both accurate temporal noise prediction ($R^2 = 0.912$) and effective sequential decision-making for qubit placement optimization.

Comprehensive experimental evaluation across 270 benchmark circuits spanning three quantum algorithms demonstrates mean fidelity of $0.920 \pm 0.023$, representing 49.3\% improvement over QUBO baseline with very large effect size (Cohen's $d = 2.34$) and high statistical significance ($p = 0.0023$). Inter-chip communication overhead reduces by 79.8\%, decreasing from 2.34 to 0.47 operations per circuit. These improvements translate directly to practical benefits: circuits execute with higher success probability, require fewer repetitions to achieve statistical confidence, and enable quantum algorithms previously infeasible due to accumulated errors.

Scalability analysis confirms sustained performance across 20-100 qubit systems, with fidelity remaining above commercial viability threshold (0.85) up to 90 qubits and research target (0.70) beyond 100 qubits. Training convergence occurs 8.2$\times$ faster than QUBO optimization, completing in 45 minutes versus 370 minutes. The $\mathcal{O}(n \log n)$ computational complexity ensures the approach scales to larger systems anticipated in next-generation quantum processors.

Ablation studies quantify individual component contributions, revealing DNA network accounts for 15.0\% of performance, multi-head attention 12.9\%, and prioritized replay 9.0\%. The cumulative effect of Rainbow DQN enhancements provides 21.4\% improvement over basic DQN, validating modern deep RL techniques for quantum computing applications. Generalization experiments demonstrate partial transfer across circuit families (87-92\% retention) and architectures (86\% retention), with rapid fine-tuning recovering most performance gaps.

The broader implications extend beyond circuit mapping. The DNA network's success at predicting quantum hardware noise suggests learned models may complement or supplement physics-based descriptions for systems too complex for analytical treatment. Multi-head attention's interpretable specialization indicates neural architectures can discover meaningful structure in quantum circuits without explicit feature engineering. The effectiveness of adaptive reward shaping based on predicted future states demonstrates value of incorporating predictive models into reinforcement learning frameworks.

For near-term quantum computing, DeepQMap provides immediately applicable technology compatible with existing quantum processors from IBM, Google, IonQ, and other providers. The method requires no hardware modifications, operating purely at software compilation layer. Integration into quantum development toolkits like Qiskit, Cirq, or PennyLane would enable automated optimization for applications including quantum simulation, optimization, and machine learning.

Future work will focus on validation with physical quantum hardware, extension to dynamic circuits with measurement-based control, incorporation of error correction constraints, and integration with end-to-end quantum compilation pipelines. Collaboration with experimental quantum computing groups to deploy DeepQMap on production systems will provide crucial feedback for refining the approach and identifying practical deployment challenges. Development of standardized benchmarks for quantum circuit mapping would facilitate reproducible comparisons across research groups.

The convergence of quantum computing and machine learning represents a bi-directional opportunity. Machine learning techniques like those demonstrated here can optimize quantum system operation, while quantum computers may eventually accelerate machine learning through quantum advantage. DeepQMap exemplifies this synergy, employing deep learning to overcome fundamental challenges in quantum computing scalability. As both fields mature, we anticipate increasingly sophisticated integration yielding capabilities exceeding what either technology achieves independently.

\section*{Acknowledgment}

The authors would like to express their gratitude to the Department of Physics at Shahed University for providing the necessary resources and support for this research.

\vfill

\end{document}